\documentclass[acmtog, authorversion]{acmart}

\usepackage{booktabs} \usepackage{wrapfig}
\usepackage{enumerate}
\usepackage{tabularx}
\usepackage{mwe}

\usepackage{bm}

\setcitestyle{square}
\setcitestyle{nosort}

\usepackage[ruled]{algorithm2e} 
\SetAlFnt{\small}
\SetAlCapFnt{\small}
\SetAlCapNameFnt{\small}
\SetAlCapHSkip{0pt}
\IncMargin{-\parindent}

\usepackage{geometrycollective}
\usepackage{VectorHeatMethod}
\usepackage{amsthm}

\acmJournal{TOG}
\acmVolume{XX}
\acmNumber{XX}
\acmArticle{00}
\acmYear{2019}
\acmMonth{6}

\setcopyright{rightsretained}

\acmDOI{0000000.0000000}

\received{May 2018}
\received[final version]{June 2018}
\received[accepted]{July 2018}

\DeclareSymbolFont{matha}{OML}{txmi}{m}{it}\DeclareMathSymbol{\varv}{\mathord}{matha}{118}

\newcommand{\R}{\mathbb{R}}

\usepackage[refpage]{nomencl}

\makenomenclature

\makeatletter
 \newcommand{\hyperraisedtarget}[1]{\Hy@raisedlink{\hypertarget{#1}{}}}
\makeatother

\newcommand{\definesymbol}[3]{\nomenclature{\(#1\)}{#2}\hyperraisedtarget{symb:#3}}

\newcommand{\createSymbol}[3]{    \ifcsname#1\endcsname                      \expandafter\renewcommand\csname #1\endcsname{{\protect\hyperlink{symb:#1}{\color{black}#2}}}    \else              \expandafter\newcommand\csname #1\endcsname{{\protect\hyperlink{symb:#1}{\color{black}#2}}}    \fi    \expandafter\newcommand\csname define#1\endcsname{\definesymbol{\csname #1\endcsname}{#3}{#1}}}

\DeclareMathOperator*{\argmin}{arg\,min}

\createSymbol{M}             {M}
{Manifold}
\createSymbol{Exp}           {\textrm{exp}}                                      {Exponential Map}
\createSymbol{Log}           {\textrm{log}}                                      {Logarithmic Map}

\begin{document}
\title{The Vector Heat Method}
\author{Nicholas Sharp}
\author{Yousuf Soliman}
\author{Keenan Crane}
\affiliation{  \institution{Carnegie Mellon University}
  \streetaddress{5000 Forbes Ave}
  \city{Pittsburgh}
  \state{PA}
  \postcode{15213}
  }

\renewcommand\shortauthors{Sharp, Soliman, and Crane}

\begin{abstract}
   This paper describes a method for efficiently computing parallel transport of tangent vectors on curved surfaces, or more generally, any vector-valued data on a curved manifold.  More precisely, it extends a vector field defined over any region to the rest of the domain via parallel transport along shortest geodesics.  This basic operation enables fast, robust algorithms for extrapolating level set velocities, inverting the exponential map, computing geometric medians and Karcher/Fr\'{e}chet means of arbitrary distributions, constructing centroidal Voronoi diagrams, and finding consistently ordered landmarks.  Rather than evaluate parallel transport by explicitly tracing geodesics, we show that it can be computed via a short-time heat flow involving the \emph{connection Laplacian}.  As a result, transport can be achieved by solving three prefactored linear systems, each akin to a standard Poisson problem.  To implement the method we need only a discrete connection Laplacian, which we describe for a variety of geometric data structures (point clouds, polygon meshes, \etc).  We also study the numerical behavior of our method, showing empirically that it converges under refinement, and augment the construction of intrinsic Delaunay triangulations (iDT) so that they can be used in the context of tangent vector field processing.
\end{abstract}

\ccsdesc[500]{Mathematics of computing~Discretization}
\ccsdesc[500]{Mathematics of computing~Partial differential equations}
\ccsdesc[500]{Computing methodologies~Shape analysis}

\keywords{discrete differential geometry, parallel transport, velocity extrapolation, logarithmic map, exponential map, Karcher mean, geometric median}

\maketitle

\begin{figure}
    \centering
    \includegraphics[width=1.0\columnwidth]{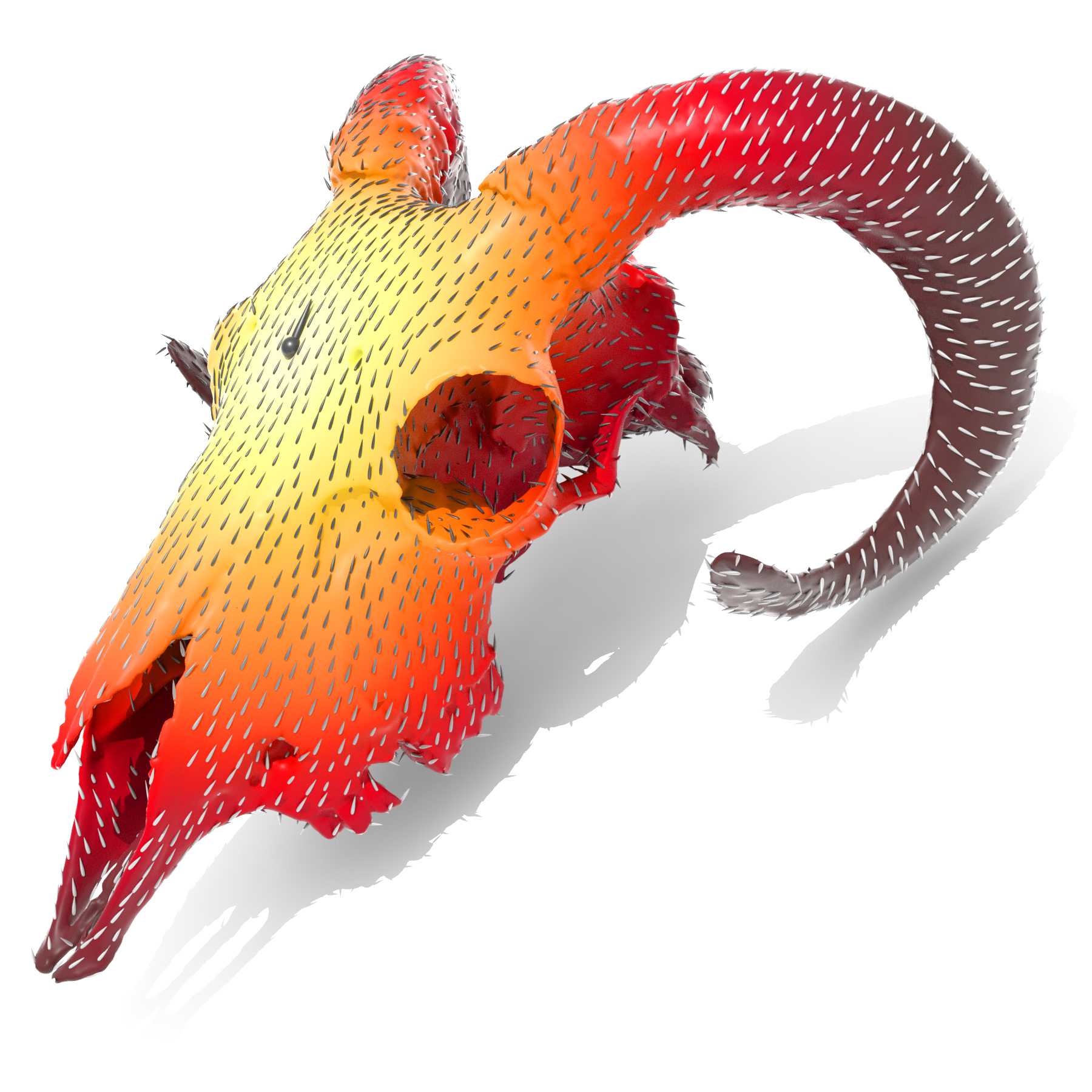}\vspace{-2\baselineskip}
    \caption{Given a vector at a point, the vector heat method computes the most parallel vector at every other point.  The method easily generalizes to other data (such as a velocity field along a curve), providing a novel and efficient way to implement fundamental algorithms across geometry and simulation.\label{fig:ram_teaser}}
\end{figure}

\section{Introduction}
\label{sec:Introduction}

Given a vector at a point of a curved domain, how do we find the most parallel vector at all other points (as shown in \figref{ram_teaser})?  This ``most parallel'' vector field---not typically considered in numerical algorithms---provides a surprisingly valuable starting point for a wide variety of tasks across geometric and scientific computing, from extrapolating level set velocity to computing centers of distributions.  To compute this field, one idea is to transport the vector along explicit paths from the source \(x\) to all other points \(y\), but even just constructing these paths is already quite expensive (\secref{RelatedWork}).  We instead leverage a little-used relationship between parallel transport and the \emph{vector heat equation}, which describes the diffusion of a given vector field over a time \(t\).  As \(t\) goes to zero, the diffused field is related to the original one via parallel transport along minimal geodesics, \ie, shortest paths along the curved domain (\secref{ConnectionLaplacian}).

The same principle applies not only to point sources, but also to vector fields over curves or other subsets of the domain.  Since diffusion equations are expressed in terms of standard Laplace-like operators, we effectively reduce parallel transport tasks to sparse linear systems that are extremely well-studied in scientific computing---and can hence immediately benefit from mature, high-performance solvers.  Moreover, since discrete Laplacians are available for a wide variety of shape representations (polygon meshes, point clouds, \etc{}), and generalize to many kinds of vector data (symmetric direction fields, differential forms, \etc{}), we can apply this same strategy to numerous applications.  In particular, this paper introduces
\begin{itemize}
   \item a fast method for computing parallel transport from a given source set (\secref{SmoothFormulation})
   \item an augmented intrinsic Delaunay algorithm for vector field processing (\secref{TangentIntrinsicDelaunay})
   \item the first method for computing a logarithmic map over the entire surface, rather than in a local patch (\secref{LogarithmicMap}), and
   \item the first method for computing true Karcher/Fr\'{e}chet means and geometric medians on general surfaces (\secref{KarcherMeans}).
\end{itemize}
We also describe how to discretize the connection Laplacian on several different geometric data structures and types of vector data (\secref{Generalizations}), and consider a variety of other applications including distance-preserving velocity extrapolation for level set methods, computing geodesic centroidal Voronoi tessellations (GCVT), and finding consistently ordered intrinsic landmarks (\secref{Applications}).

\section{Related Work}
\label{sec:RelatedWork}

\paragraph{Discrete Parallel Transport} Parallel transport has a long history in the discrete setting.  One of the earliest ideas, perhaps, is \emph{Schild's ladder} which approximates parallel transport via short geodesic segments; this technique has proven useful for parallel transport in high-dimensional spaces representing image data~\cite{Lorenzi:2014:EPT}, but is not directly related to parallel transport of vectors on discrete surfaces.  A more natural predecessor to the type of parallel transport encountered in geometry processing is the simplicial calculus of \citet{Regge:1961:GRC}, largely used for problems in general relativity~\cite{Gentle:2002:RCU}.  On surfaces, this approach essentially amounts to the notion of \emph{discrete connections} studied by \citet{Crane:2010:TCD}.  To discretize our method on triangle meshes, we will instead consider vectors at vertices, building on ideas from \citet{Polthier:1998:SGP} and \citet{Knoppel:2013:GOD}.  Finally, \citet{Azencot:2015:DDV} explore a spectral approach to parallel transport, though here the goal is different from ours: transporting one vector field along another, rather than transporting vectors along shortest geodesics.

\setlength{\columnsep}{1em}
\setlength{\intextsep}{0em}
\begin{wrapfigure}{r}{1.7in}
   \includegraphics[width=1.7in]{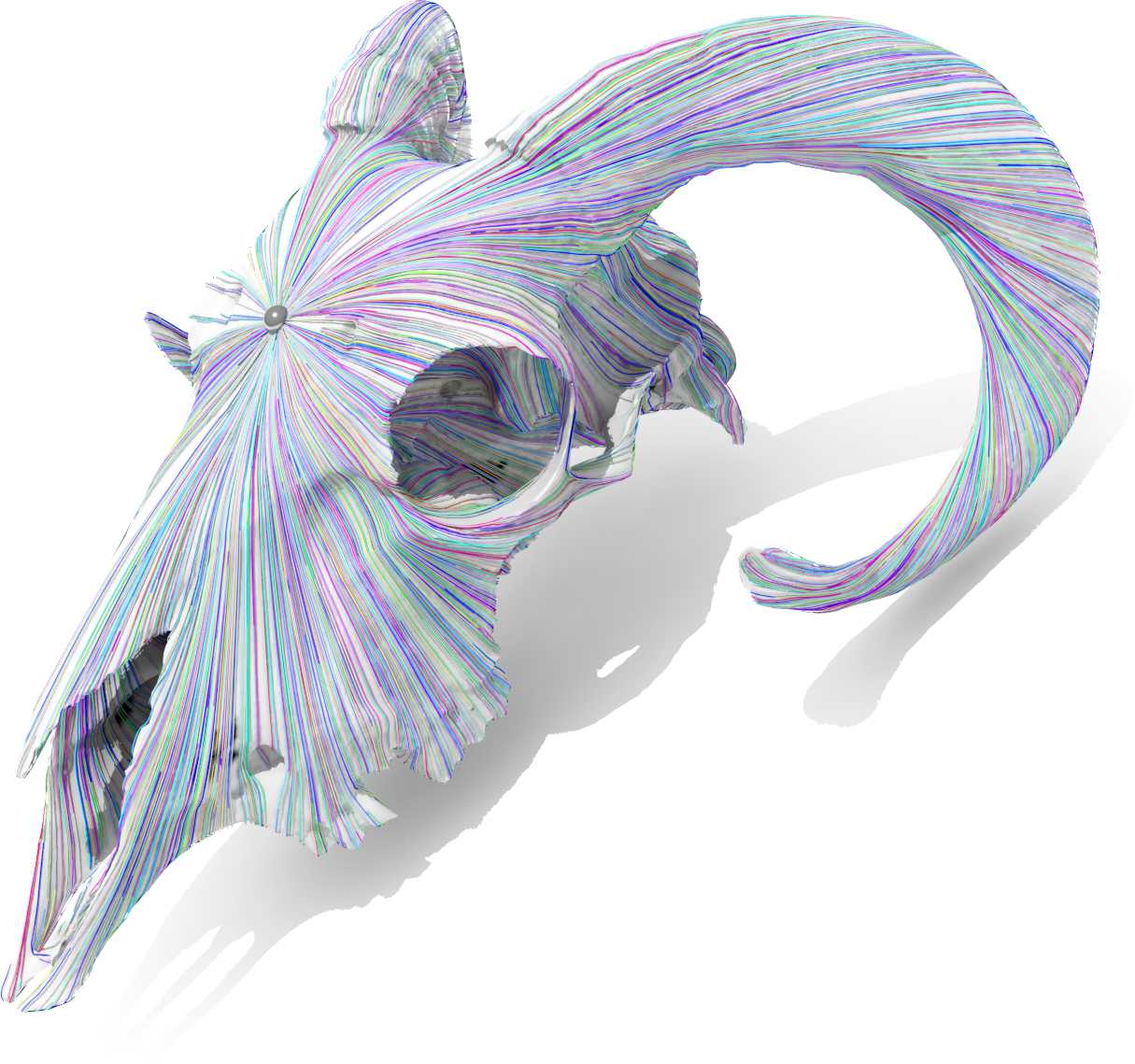}
   \caption{Even just \emph{tracing} all the geodesics to a given source point (without accounting for the cost of distance computation) is already an order of magnitude more expensive than applying the vector heat method---here, 15x slower.\label{fig:curve_tracing}}
\end{wrapfigure}

\paragraph{Discrete Geodesics} A seemingly natural solution to our problem is to explicitly transport vectors along geodesic paths---in the case of triangle meshes, one could unfold the triangles along the path and apply a simple translation in the plane (\ala{} \citet{Polthier:1998:SGP}).  However, finding these paths is not straightforward: one can either compute exact polyhedral geodesics via expensive window-based methods \cite{Mitchell:1987:DGP,Chen:1990:SPP} that demand sophisticated acceleration schemes \cite{Surazhsky:2005:FEA,Ying:2013:SVG,Qin:2016:FED}; or trace integral curves of a piecewise linear geodesic distance function \cite{Kimmel:1998:FMM,Crane:2013:GHN}, which may have very different behavior from true geodesics \cite{Tricoche:2000:HOS}.  Our approach is far simpler: just build Laplace matrices and solve linear systems.  It is also more efficient: even if there were no cost associated with computing geodesics, each shortest path on a discrete surface with \(O(n)\) elements has length \(O(\sqrt{n})\), yielding an overall cost in \(O(n^{1.5})\).  In our method, the cost is dominated by solving sparse diffusion equations, which has complexity approaching \(O(n)\) for both iterative and direct methods~\cite{Spielman:2004:NTA,Gillman:2014:DSO}; prefactorization can be used to further reduce amortized cost across many different source points or sets (\secref{ImplementationAndPerformance}).  In practice we observe that merely \emph{extracting} paths from a given piecewise constant vector field is more than an order of magnitude slower than executing our entire algorithm (\figref{curve_tracing}).  Moreover, the diffusion-based approach also provides an accurate and reliable solution (\secref{ConvergenceAndAccuracy}).

\paragraph{Relationship to Scalar Heat Method} The original, scalar heat method \cite{Crane:2013:GHN} computes a related, but fundamentally different quantity from the vector heat method: the former computes geodesic distance; the latter computes parallel transport along shortest geodesics.  Computationally, these methods share some basic features: rather than directly solve a difficult nonlinear hyperbolic problem (wavefront propagation from a source), they reformulate computation in terms of much easier linear elliptic PDEs (local averaging); all nonlinearity is captured by simple pointwise operations.  However, the structure of the vector version is different: unlike the scalar heat method, there is no dependence among linear equations (\StepI{}--\StepIII{} of \algref{vector_heat_method}), making error behavior easier to analyze, and
providing additional opportunities for acceleration.  Moreover, the vector heat method does not require discrete divergence or gradient operators, making it easier to apply to data structures like point clouds, or even (in principle) data on general graphs~\cite{Karoui:2013:GCL}.

\paragraph{Connection Laplacians} On flat domains like the plane, vector diffusion amounts to diffusion of individual scalar components.  On curved domains things are not so simple: there is typically no global coordinate system, and one must therefore apply a diffusion process that accounts for parallel transport, achieved via the \emph{connection Laplacian} (\secref{Preliminaries}). \citet{Singer:2012:VDM} use a similar process to obtain a \emph{vector diffusion distance}, motivated by tasks in data analysis and machine learning.  \citet{Lin:2014:GDF} likewise consider vector diffusion in the learning context; we leverage a similar technique in \algref{log_map}, \StepII{}, deriving initial conditions that substantially improve accuracy.  On triangle meshes, \citet{Knoppel:2013:GOD,Knoppel:2015:SPS} consider two connection Laplacians: one based on finite elements, and another in the spirit of \emph{discrete exterior calculus}~\cite{Desbrun:2006:DDF}; we build primarily on the latter.  Algorithmically, fast solvers for connection Laplacians are an active area of research~\cite{Kyng:2016:SCM}; applications built on top of the vector heat method can immediately benefit from new developments in this area.

\paragraph{Applications} Though we postpone detailed background on applications to \secref{Applications}, it is worth noting that the heat flow approach is the first practical way to compute an accurate \emph{logarithmic map} (sometimes referred to by its inverse, the \emph{exponential map}) over the entire domain rather than just a local patch---\figref{octopus_salad} provides a comparison with previous methods, which exhibit significant error over longer distances.  This global accuracy in turn yields the first efficient and reliable method for computing \emph{Karcher means} and \emph{geometric medians} on arbitrary surfaces (see especially \figref{KarcherMeanComparison}).  Computationally, previous methods are Dijkstra-like and necessitate dynamic branches and different memory access patterns for each source point.  In contrast, heat methods execute a fixed and hence highly predictable traversal of a minimal data structure (a matrix factorization).  As a result, the constants involved tend to be much smaller---for instance, our log map computation is faster than even the basic method of \citet{Schmidt:2006:IDC}.  More broadly, tasks that depend on global integration of information (such as computing means or landmarks) benefit from the robust global nature of our algorithm.

\section{Preliminaries}
\label{sec:Preliminaries}

The basic idea of our method is to approximate parallel transport via short-time diffusion of vector-valued data.  In the Euclidean setting, one can simply diffuse individual scalar components via the ordinary heat equation, but on curved domains this approach fails, since for vectors in different tangent spaces equality of coordinates has no geometric significance.  We instead consider a particular vector heat equation which, for a short time \(t\), keeps vectors parallel.  We first provide some basic notation and definitions.

\begin{figure}[t]
   \includegraphics{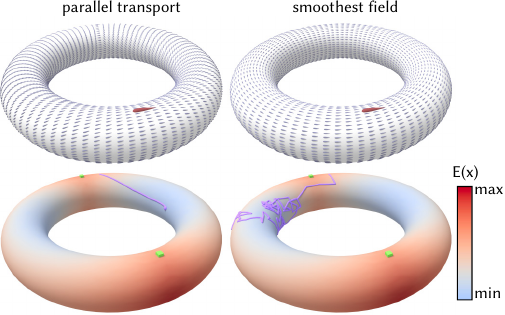}
   \caption{The field obtained by parallel transport along shortest geodesics plays a special role in several basic geometric algorithms---replacing it with a field that is merely \emph{smooth} can cause these algorithms to fail.  Here for instance we use the field to locate the \emph{Karcher mean} equidistant from the two green points \figloc{(left)}.  Attempting to instead use the smoothest possible field causes this search to fail, wandering randomly over the surface \figloc{(right)}.\label{fig:smoothest_field_comparison}}
\end{figure}

\subsection{Notation}
\label{sec:Notation}

Throughout we consider a Riemannian manifold \(M\) with metric \(g\). We use \(d(x,y)\) to denote the corresponding geodesic distance, \ie, the length of the shortest path between any two points \(x,y \in M\).  The \emph{cut locus} of any subset \(\Omega \subseteq M\) is the set of all points \(p \in M\) for which there is not a unique closest point \(q \in \Omega\) (\figref{schmidt_comparison_set}, \figloc{bottom}).  For a vector field \(X\) on \(M\), we use \(X|_p\) to denote the vector at a point \(p\).  We will use \(\imath \in \CC\) to denote the imaginary unit, \ie, \(\imath^2 = -1\).  Finally, we use \(\delta_x\) to denote the Dirac delta centered at \(x \in M\).

\vspace{2\baselineskip}

\subsection{Heat Diffusion}
\label{sec:Heat Diffusion}

The most basic diffusion equation is the scalar \emph{heat equation}, which describes how an initial heat distribution \(\phi_0: M \to \RR\) looks after being diffused for a time \(t > 0\):
\begin{equation}
   \label{eq:scalar_heat_equation}
   \tfrac{d}{dt} \phi_t = \Delta \phi_t.
\end{equation}
The operator \(\Delta\) is the (negative semidefinite) Laplace-Beltrami operator on \(M\); in Euclidean \(\RR^n\), \(\Delta\) is just the usual Laplacian.

\paragraph{Heat Kernel} When the initial heat distribution \(\phi_0\) is just a spike \(\delta_x\) at a single point \(x\), the solution to \eqref{scalar_heat_equation} is referred to as the \emph{heat kernel} \(k_t\).  The heat kernel is the \emph{fundamental solution}, in the sense that convolution of the initial data \(\phi_0\) with \(k_t\) yields the solution to the heat equation at time \(t\).  When the domain is Euclidean (\ie, \(M = \RR^n\)), this fundamental solution is just a Gaussian of constant total mass, centered at a point \(x \in \RR^n\):
\begin{equation}
   \label{eq:gaussian_kernel}
   G_t(x,y) := \frac{1}{(4\pi t)^{n/2}} e^{-d(x,y)^2/4t}.
\end{equation}
Here \(n\) denotes the dimension of the domain (\eg, \(n=2\) for the plane).  More generally, the heat kernel has the asymptotic expansion 
\begin{equation}
   \label{eq:scalar_heat_kernel}
   k_t(x,y) \sim \frac{e^{-d(x,y)^2/4t}}{(4\pi t)^{n/2}} j(x,y)^{-1/2} \left( 1 + \sum_{i=1}^\infty t^i \Phi_i(x,y) \right).
\end{equation}
For our purposes the definition of the functions \(j\) and \(\Phi_i\) will not be important, especially since we consider the limit as \(t \to 0\) (see \cite[Theorem 2.30]{Berline:1992:HKD} for further discussion). In practice, we obtain a numerical approximation of \(k_t\) by solving \eqref{scalar_heat_equation} directly, \ie, by placing a Dirac delta at a source point \(x\) and ``smearing'' it out via heat diffusion.

\subsection{Parallel Transport and Connections}
\label{sec:ParallelTransportAndConnections}

\setlength{\columnsep}{1em}
\setlength{\intextsep}{0em}
\begin{wrapfigure}{r}{90pt}
   \includegraphics{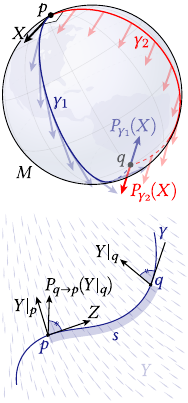}
\end{wrapfigure}
Given a tangent vector \(X\) at a point \(p\) of a curved surface \(M\), which vector at another point \(q\) should be considered ``parallel?''  If we have a smooth curve \(\gamma(s)\) going from \(p\) to \(q\), one reasonable idea is that \(X\) should experience no ``unnecessary turning,'' \ie, no change along the tangent direction \(T := \tfrac{d}{ds}\gamma\); the vector we obtain at the end of the path is called the \emph{parallel transport} of \(X\) along \(\gamma\), which we will denote by \(P_{\gamma}(X)\).  An important fact about parallel transport is that it is \emph{path dependent}, \ie, for two different curves \(\gamma_1,\gamma_2\) from \(p\) to \(q\), it is not necessarily true that \(P_{\gamma_1}(X) = P_{\gamma_2}(X)\).  A good example is transporting a vector from the north to the south pole of the Earth along two different lines of longitude: at the south pole, the angle between the resulting vectors will be related to the difference in longitudes (see inset, \figloc{top}).  However, as \(q\) gets closer and closer to \(p\), only the outgoing direction of the path matters, since very short segments of paths with the same tangent become indistinguishable.  We can therefore use parallel transport to define the derivative of one vector field \(Y\) along another vector field \(Z\).  In particular, at any point \(p \in M\) the \emph{covariant derivative} \(\nabla_Z Y\), describes the change in \(Y\) as we travel an infinitesimally short distance along \emph{any} curve \(\gamma\) with tangent \(Z\) at \(p\).  More formally, letting \(p = \gamma(0)\) and \(q = \gamma(s)\), \(\nabla_Z Y|_p := \lim_{s \to 0} (P_{q \to p} (Y|_q) - Y|_p)/s\), where \(P_{q \to p}\) denotes parallel transport from \(q\) back to \(p\) (see inset, \figloc{bottom}).  The operator \(\nabla\) is referred to as the \emph{Levi-Civita connection}.

\begin{figure}[b]
   \includegraphics{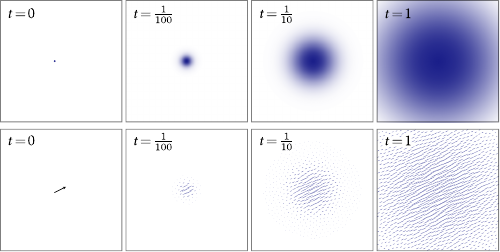}
   \caption{Similar to the scalar heat kernel \(k_t\) \figloc{(top)}, the vector heat kernel \(k^\nabla_t\) \figloc{(bottom)} ``smears out'' vectors over time.  In the flat Euclidean case one can simply diffuse each scalar component independently, but on a curved domain the \emph{connection Laplacian} is needed to diffuse vectors from one tangent space to another. (Figures not to scale.)\label{fig:vector_diffusion}}
\end{figure}

\subsection{Connection Laplacian}
\label{sec:ConnectionLaplacian}

The \emph{connection Laplacian} \(\ConnectionLaplacian\) is a second derivative on vector fields with many of the same basic properties as the ordinary Laplacian~\(\Delta\): it is negative semidefinite, self-adjoint, and elliptic.  Just as the ordinary negative semidefinite Laplacian can be expressed as the trace of the Hessian, or as the divergence of the gradient (\(\Delta f = \mathrm{tr}(H(f)) = \mathrm{div} \circ \mathrm{grad} f\)), the connection Laplacian associated with a connection \(\nabla\) is given by the trace of the second covariant derivative, or by the composition of the covariant derivative with its adjoint (\(\ConnectionLaplacian X = \mathrm{tr}(\nabla^2 X) = -\nabla^\ast \nabla X\)).  Some intuition can be obtained by relating the connection Laplacian to the \emph{vector heat equation}
\begin{equation}
   \label{eq:vector_heat_equation}
   \tfrac{d}{dt} X_t = \ConnectionLaplacian X_t.
\end{equation}
Intuitively, the evolution of the vector field \(X_t\) over time will look like a ``smearing out'' of an initial vector field \(X\) (\figref{vector_diffusion}).  We can make this statement more precise by considering the associated heat kernel \(k^\nabla_t(x,y)\), which describes how a vector at a single point \(x\) will diffuse to all other points \(y\) over time \(t\).  For points \(y\) that are not on the cut locus of \(x\), this kernel has the asymptotic expansion 
\begin{equation}
   \label{eq:vector_heat_kernel}
   k^\nabla_t(x,y) \sim \frac{e^{-d(x,y)^2/4t}}{(4\pi t)^{n/2}} j(x,y)^{-1/2} \left( \sum_{i=0}^\infty t^i \Psi_i(x,y) \right),
\end{equation}
where the functions \(\Phi_i\) from the scalar heat kernel have been replaced by maps \(\Psi_i\) taking vectors at \(x\) to vectors at \(y\).  Most importantly, the first function in this series is given by
\begin{equation}
   \label{eq:lowest_order_term}
   \boxed{\Psi_0(x,y) = P_{\gamma_{x \to y}},}
\end{equation}
where \(\gamma_{x \to y}\) is the shortest curve from \(x\) to \(y\), \ie, the \emph{shortest geodesic}~\cite[Theorem 2.30]{Berline:1992:HKD}.  In other words, as \(t \to 0\), the vector heat kernel behaves like parallel transport along shortest paths, along with a decay in magnitude that is identical to the decay of the scalar heat kernel. (As a side note: for \(t \to \infty\), \(X_t\) approaches the smoothest possible vector field---independent of initial conditions---since the vector heat equation corresponds to gradient descent on the vector Dirichlet energy; normalizing this field would yield the optimal direction field in the sense of \citet{Knoppel:2013:GOD}.)

Note that not all vector diffusion equations yield the same behavior: for instance, a vector diffusion equation formulated in terms of the \emph{Hodge Laplace operator} (discussed in \secref{Differential1Forms}) will exhibit different behavior with respect to parallel transport.  The discrete picture also provides some useful intuition for the connection Laplacian---see \secref{DiscreteLaplaceOperators}.

\begin{figure}[t]
   \includegraphics[width=\columnwidth]{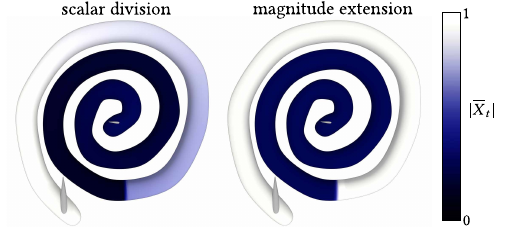}
   \caption{Numerically, just taking the quotient of short-time solutions to the vector and scalar heat equations yields a poor approximation of magnitude \figloc{(left)}, which in this case should be piecewise constant.  By separately computing magnitudes (\figloc{right}), we obtain a more accurate solution.\label{fig:quotient_vs_maginterpolate}}
\end{figure}

\section{Smooth Formulation}
\label{sec:SmoothFormulation}

The relationship between the vector heat kernel and the parallel transport map (\eqref{lowest_order_term}) is critical to our method, since it allows us to compute parallel transport by solving diffusion problems---which in turn amount to easy linear systems.  For instance, to transport a single unit vector to the rest of the surface, one could simply compute the vector heat kernel for small time~\(t\), then normalize the resulting vectors.  In the general case, this strategy will not work: consider three vectors of different magnitudes, or a vector field of varying magnitude along a curve (\figref{bean_vector_combined}).  One way to account for this varying magnitude is to observe that the scalar heat kernel \(k_t\) (\eqref{scalar_heat_kernel}) and the vector heat kernel \(k^\nabla_t\) (\eqref{vector_heat_kernel}) have identical leading coefficients.  Therefore, as \(t \to 0\) the higher-order terms vanish and we can recover the parallel transport map as a simple quotient:
\begin{equation}
   \label{eq:HeatKernelRatio}
   \lim_{t \to 0} \frac{k^\nabla_t(x,y)}{k_t(x,y)} = P_{\gamma_{x \to y}}.
\end{equation}
More generally, suppose we diffuse a given vector field \(X\) supported (\ie, nonzero) on a set \(\Omega\), and diffuse the corresponding scalar indicator function \(\phi_0 = \Indicator_\Omega\) (formally, a Hausdorff measure of appropriate dimension).  Since diffusion is equivalent to convolution with the heat kernel, these quantities approach the same magnitude at each point, \ie,
\[
   \lim_{t \to 0} |X_t| - \phi_t = 0.
\]
Hence, the quotient \(X_t/\phi_t\) should exactly factor out any decay in magnitude, leaving only the result of parallel transport along shortest geodesics.  Numerically, however, the situation is not so simple: even for fairly small values of \(t\), diffused vectors pointing in different directions will yield small cancellation errors, further reducing the magnitude of the numerator \(X_t\) (\figref{quotient_vs_maginterpolate}, \figloc{left}).  To get reliable numerical results we will need to consider an alternative approach: use scalar diffusion to obtain the magnitude of the transported vectors (\secref{ScalarInterpolation}); use vector diffusion to obtain their direction (\secref{VectorHeatMethod}).  Together these operations define our basic algorithm (\algref{vector_heat_method}), though nothing restricts the method to two dimensional surfaces, nor to the tangent bundle: everything we state in the smooth setting immediately applies to any vector bundle over a Riemannian manifold of any dimension, as we will discuss in \secref{Generalizations}.

\subsection{Scalar Interpolation}
\label{sec:ScalarInterpolation}

\begin{figure}
   \centering
   \includegraphics{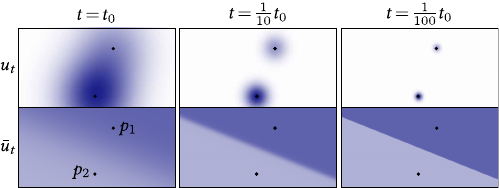}
   \caption{A simple example of interpolation by diffusion: a Gaussian weighted combination of values at two points \(p_1,p_2\), yields a closest point interpolation \(\overline{u}_t\) as the Gaussian width \(t\) goes to zero.  Here \(u_t\) is the linear combination prior to normalization.\label{fig:plane_example}}
\end{figure}

Suppose we have a pair of source points \(p_1, p_2 \in \RR^2\), with associated values \(u_1, u_2 \in \RR\).  How can we find a function \(\overline{u}\) over the rest of the plane whose value at each point is equal to the value at the closest source \(p_i\)?  For this particular example the answer is obvious (just find the line separating \(p_1\) and \(p_2\)), but we can obtain it in an interesting way that will naturally generalize.  Suppose we use the Gaussian kernel (\eqref{gaussian_kernel}) to define a weighted average
\[
   \overline{u}_t = \frac{u_1 G_{t,p_1} + u_2 G_{t,p_2}}{G_{t,p_1} + G_{t,p_2}}.
\]
As \(t\) goes to zero, this weighted average provides a closest point interpolation (\figref{plane_example}, \figloc{bottom}), since for points closer to \(p_1\) than \(p_2\), the numerator \(u_t := u_1 G_{p_1,t} + u_2 G_{p_2,t}\) is dominated by the first term, and vice versa (\figref{plane_example}, \figloc{top}).

This basic idea is easily generalized to curved domains: interpolation is again achieved by dividing a weighted sum by the sum of weights, except that we replace the Gaussian kernel \(G_t\) with the scalar heat kernel \(k_t\) (\eqref{scalar_heat_kernel}).  In particular, given a collection of sources \(p_1, \ldots, p_n \in M\) and associated values \(u_1, \ldots, u_n \in \RR\), we solve two independent heat equations for functions \(u\) and \(\phi\), using initial conditions
\[
   \begin{array}{rcl}
      u_0    &=& \sum_{i=1}^n u_i \delta_{p_i}, \\
      \phi_0 &=& \sum_{i=1}^n \delta_{p_i}.
   \end{array}
\]
The interpolant is then simply the limit as \(t\) goes to zero of the normalized function
\[
   \overline{u}(t) := \frac{u(t)}{\phi(t)}.
\]
The intuition is the same as in the planar case: for points closest to \(p_i\), the weighted sum will be dominated by the \(u_i\) term.  Points exactly on the cut locus will approach an average of values; though a precise analysis of this behavior becomes more difficult~\cite{Grigoryan:2009:HKA}, in practice these values are well-behaved.

\begin{figure*}
   \begin{minipage}{.47\textwidth}
      \includegraphics{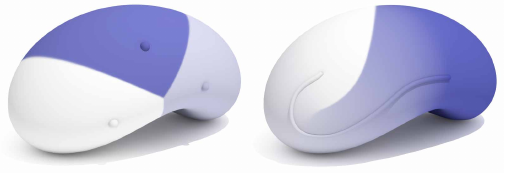}
   \caption{Interpolation of scalar values at points \figloc{(left)} and curves \figloc{(right)}. \label{fig:scalar_examples}}
   \end{minipage}
   \hspace{.05\textwidth}
   \begin{minipage}{.47\textwidth}
      \includegraphics{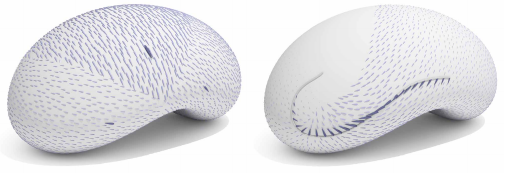}
   \caption{Parallel transport of vectors from points \figloc{(left)} and curves \figloc{(right)}. \label{fig:bean_vector_combined}}
   \end{minipage}
\end{figure*}

More generally, the source set can be any subset \(\Omega \subset M\)---on a surface, for instance, \(\Omega\) can be a collection of points, curves, and regions (see for instance \figref{scalar_examples}).  In this case, the initial conditions are essentially a Dirac-type measure concentrated on \(\Omega\) (or more formally, a sum of Hausdorff measures of appropriate dimension); in the discrete setting we can integrate basis functions with respect to this measure to obtain initial conditions (as in \appref{DistanceGradientDiscretization}).

\begin{algorithm}
\caption{Vector Heat Method}
\label{eq:fpm}
   \textbf{Input:} A vector field \(X\) supported on a subset \(\Omega \subset M\) of the domain \(M\). \\
   \textbf{Output:} A vector field \(\overline{X}\) on all of \(M\). \\
\begin{enumerate}[I.]
   \item Integrate the vector heat flow \(\tfrac{d}{dt} Y_t = \ConnectionLaplacian Y_t\) for time \(t\), with \(Y_0 = X\).
   \item Integrate the scalar heat flow \(\tfrac{d}{dt} u_t = \Delta u_t\) for time \(t\), with \(u_0 = |X|\).
   \item Integrate the scalar heat flow \(\tfrac{d}{dt} \phi_t = \Delta \phi_t\) for time \(t\), with \(\phi_0 = \Indicator_\Omega\).
   \item Evaluate the vector field \(\overline{X}_t = u_t Y_t / \phi_t |Y_t|\).
\end{enumerate}
\label{alg:vector_heat_method}
\end{algorithm}

\subsection{Vector Heat Method}
\label{sec:VectorHeatMethod}

We now define our main algorithm, the \textbf{vector heat method}, which is summarized in \algref{vector_heat_method}.  The basic idea is to first diffuse a given vector field via the vector heat equation (\eqref{vector_heat_equation}).  For small time \(t\), the resulting vectors will have essentially the right direction, but the wrong magnitude.  To obtain the right magnitude, we interpolate the magnitudes of the source vectors (as in \secref{ScalarInterpolation}), and scale the normalized vectors by these magnitudes.  The result is a field where the vector at each point \(q\) closely approximates the parallel vector at the closest point \(p\).  More precisely, for any given vector field \(X\) supported on a subset \(\Omega \subset M\) of the domain \(M\), we obtain a vector field \(\overline{X}_t\) such that at each point \(q \in M\) not in the cut locus of \(\Omega\),
\[
   \lim_{t \to 0} \overline{X}_t|_q = P_{\gamma_{p \to q}} X|_p,
\]
where \(p\) is the point of \(\Omega\) closest to \(q\), and \(\gamma_{p \to q}\) is the shortest geodesic from \(p\) to \(q\).  In practice we cannot evaluate the limit field directly; instead, we pick a small time step \(t > 0\) (as detailed in \secref{ConvergenceAndAccuracy}) and solve the vector diffusion equation
\[
   \begin{array}{rclll}
      \tfrac{d}{dt} Y_t  &=& \ConnectionLaplacian Y_t  &\text{with}& Y_0 = X, \\
   \end{array}
\]
to obtain a diffused vector field \(Y_t\).  For a small time \(t > 0\), each vector \(Y_t|_q\) will closely match the \emph{direction} of the closest source vector \(X|_p\), but will have the wrong \emph{magnitude}, \ie, \(|Y_t|_q \ne |X|_p\).  To get the right magnitude, we solve the two scalar diffusion equations
\[
   \begin{array}{rclll}
      \tfrac{d}{dt} u_t     &=& \Delta u_t     &\text{with}&    u_0 = |X|, \qquad \text{and} \\
      \tfrac{d}{dt} \phi_t  &=& \Delta \phi_t  &\text{with}& \phi_0 = \Indicator_\Omega. \\
   \end{array}
\]
The quotient \(\overline{u}_t := u_t/\phi_t\) then gives the magnitude of the vector at the closest point, and the final vector field is hence just
\[
   \overline{X}_t = \overline{u}_t Y_t/|Y_t|.
\]
This strategy resembles the scalar heat method, where one cannot simply apply Varadhan's formula, but must instead normalize the gradient to obtain the correct magnitude.  Likewise, in the vector heat method we cannot simply divide the vector heat kernel by the scalar heat kernel (as in \eqref{HeatKernelRatio}), but must carefully interpolate magnitudes.  This strategy provides numerical robustness: even if there are small errors in direction, the magnitudes are essentially perfect (\figref{quotient_vs_maginterpolate}, \figloc{right}).

Note that for points on the cut locus, \(\overline{X}_t\) essentially approaches an average of all closest vectors (see for instance \citet{Ludewig:2016:HKA}).  Importantly, we have no particular interest in approximating the cut locus itself; the presence of a cut locus is merely a natural feature of any globally accurate approximation of the true (smooth) solution to our problem.  Global accuracy turns out to be essential for applications---see for instance \figref{smoothest_field_comparison} and discussion in \secref{Applications}.

\section{Discrete Formulation}
\label{sec:DiscreteFormulation}

Fundamentally, the vector heat method is an algorithm formulated in the smooth setting---so far we have not assumed that we work with any particular discretization (such as point clouds or polygon meshes).  In this section we discretize the method on triangle meshes; other possibilities are explored in \secref{OtherDiscretizations}.

\subsection{Discrete Surface}
\label{sec:DiscreteSurface}

\paragraph{Topology} Throughout we consider a manifold triangle mesh \(K = (V,E,F)\), with or without boundary.  In principle our method applies to nonorientable domains, but to simplify exposition (and implementation) it will be easier to assume that \(K\) is oriented.  We use tuples of vertex indices to specify simplices---for instance, \(ijk\) is a triangle with vertices \(i,j,k \in V\).  Indices appearing on both sides of an equation are held fixed in sums, for instance, \(a_{ij} = \sum_{ijk \in F} b_{ijk}\) denotes a sum over only those triangles \(ijk \in F\) containing edge \(ij\).

\paragraph{Geometry} The only geometric information we need to formulate our algorithm is positive edge lengths \(\ell: E \to \RR_{>0}\) satisfying the triangle inequality in each face; from this data one can easily determine the area \(A_{ijk}\) of each triangle, and the interior angle \(\smash{\theta_i^{jk}}\) at each corner \(i\) of each triangle \(ijk\) (via Heron's formula and the law of cosines, \resp{}).  For problems involving tangent vector fields, this purely \emph{intrinsic} point of view has some attractive consequences---in particular, it enables us to talk about tangent vector fields on an \emph{intrinsic Delaunay triangulation} (\secref{TangentIntrinsicDelaunay}), which in practice can significantly improve accuracy and reliability (\figref{intrinsic_delaunay_maps}).

\setlength{\columnsep}{1em}
\setlength{\intextsep}{0em}
\begin{wrapfigure}{r}{110pt}
   \vspace{-\baselineskip}
   \includegraphics{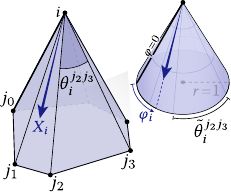}
\end{wrapfigure}
\subsection{Intrinsic Tangent Spaces}
\label{sec:IntrinsicTangentSpaces}

At each vertex \(i \in V\), we encode tangent vectors \(X_i\) in local polar coordinates \((r_i,\varphi_i)\), \ala{} \citet{Knoppel:2013:GOD}.  Conceptually, one can imagine isometrically mapping a small neighborhood of the vertex onto a circular cone whose base has a radius \(r=1\) (see inset); the direction of any tangent vector can then be expressed as an angle \(\varphi \in [0,2\pi)\), equal to the arc length along the cone boundary.  Concretely, we pick a canonical reference edge \(ij_0\) to represent the direction \(\varphi=0\); all other directions are expressed as a counter-clockwise rotation relative to this edge.  In particular, letting
\[
   \Theta_i := \sum_{ijk \in F} \theta_i^{jk}
\]
be the total interior angle at vertex \(i\), we define normalized angles
\begin{equation}
   \label{eq:normalized_angles}
   \tilde{\theta}_i^{jk} := 2\pi\theta_i^{jk}/\Theta_i,
\end{equation}
which sum to \(2\pi\).  The direction of the outgoing edges \(ij_0, ij_1, \ldots\) (in counter-clockwise order) are then given by the cumulative sums
\begin{equation}
   \label{eq:outgoing_edge_angles}
   \varphi_{ij_a} := \sum_{p=0}^{a-1} \tilde{\theta}_i^{j_p,j_{p+1}}.
\end{equation}
A tangent vector \(X_i\) at any vertex \(i\) is specified by an angle \(\varphi_i\) and magnitude \(r_i\) in this coordinate system.  In practice, we will encode this data as a complex number \(r_i \ee^{\imath\varphi_i} \in \CC\).

\setlength{\columnsep}{1em}
\setlength{\intextsep}{0em}
\begin{wrapfigure}{r}{75pt}
   \includegraphics{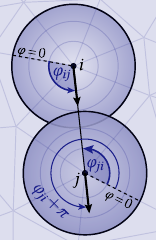}
\end{wrapfigure}
\paragraph{Discrete Parallel Transport} For any edge \(ij \in E\), the angles \(\varphi_{ij}\) and \(\pi + \varphi_{ji}\) encode the edge direction relative to the coordinate systems at vertices \(i\) and \(j\), \resp{}  To keep a vector parallel as we go from \(i\) to \(j\), we must therefore rotate by the angle
\[
   \hspace{-70pt}\rho_{ij} := (\varphi_{ji} + \pi) - \varphi_{ij}.
\]
We encode the corresponding rotations as unit complex numbers
\begin{minipage}[t][2\baselineskip][c]{75pt} 
\begin{equation}
   \label{eq:parallel_transport_rotations}
   \hspace{-70pt}r_{ij} := e^{\imath\rho_{ij}},
\end{equation}
\end{minipage}
which will help to construct our discrete connection Laplacian.
\subsection{Discrete Laplace Operators}
\label{sec:DiscreteLaplaceOperators}

For triangle meshes, the Laplace-Beltrami operator \(\Delta\) can be discretized as a weighted graph Laplacian \(\Lsf \in \RR^{|V| \times |V|}\), given by
   \vspace{.5\baselineskip}
\[
   (\Lsf\phi)_i = \tfrac{1}{2} \sum_{ij \in E} \underbrace{(\cot\theta_k^{ij} + \cot\theta_l^{ji})}_{=:w_{ij}}( \phi_j - \phi_i )
\]
\setlength{\columnsep}{1em}
\setlength{\intextsep}{0em}
\begin{wrapfigure}{r}{48pt}
   \includegraphics{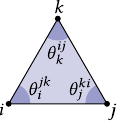}
\end{wrapfigure}
at each vertex \(i \in V\), where \(k,l\) denote the vertices opposite edge \(ij\); the \emph{cotan weights} \(w_{ij}\) simply account for the shape of the triangles (see \citet[Chapter 6]{Crane:2013:DGP}).  To be concrete, let \(a := \cot\theta_i^{jk}\), \(b := \cot\theta_j^{ki}\) and \(c := \cot\theta_k^{ij}\) be the cotangents of the angles at the corners of a triangle \(ijk \in F\).  One way to build \(\Lsf\) is to accumulate, for each triangle, the local \(3 \times 3\) matrix
\[
  -\frac{1}{2}\left[
      \begin{array}{ccc}
        b + c &  -c     &  -b  \\
         -c   & c + a   &  -a  \\
         -b   &  -a     & a + b
      \end{array}
   \right]
\]
into the corresponding entries of \(\Lsf\).

\begin{figure}[b]
   \includegraphics{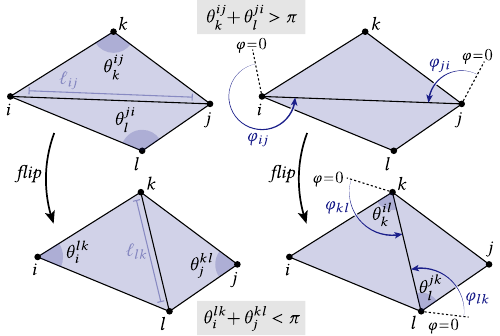}
   \caption{\figloc{Left:} the intrinsic Delaunay algorithm performs edge flips to obtain a mesh where edge angle sums are no greater than \(\pi\).  \figloc{Right:} we augment this algorithm to update the local tangent spaces on each flip.\label{fig:delaunay_flip}}
\end{figure}

The connection Laplacian \(\ConnectionLaplacian\) is given by a nearly identical \emph{complex} matrix \(\LsfNabla \in \CC^{|V| \times |V|}\); the only change is that the off-diagonal entries are multiplied by the edge rotations \(r_{ij}\):
\[
  -\frac{1}{2}\left[
      \begin{array}{ccc}
        b + c       &  -c r_{ij} &  -b r_{ik}  \\
         -c  r_{ji} & c + a      &  -a r_{jk} \\
         -b  r_{ki} &  -a r_{kj} & a + b
      \end{array}
   \right].
\]
This matrix is Hermitian since \(r_{ij}\) and \(r_{ji}\) are unit complex numbers encoding equal and opposite rotations; hence, \smash{\(r_{ji} = r_{ij}^{-1} = \overline{r}_{ij}\)}. This matrix naturally arises as the Hessian of the vector Dirichlet energy \smash{\(\sum_{ij \in E} w_{ij} |X_j - r_{ij}X_i|^2\)}, which quantifies the ``straightness'' of a given vector field \(X\) (see \citet[Section 3.2]{Knoppel:2015:SPS}).  Both \(\Lsf\) and \(\Lsf^\nabla\) effectively encode zero Neumann boundary conditions; zero Dirichlet conditions will yield similar results (see discussion in \citet[Section 3.4]{Crane:2013:GHN}).  We also have a diagonal \(|V| \times |V|\) \emph{lumped mass matrix} with entries
\[
   \Msf_{ii} = \tfrac{1}{3} \sum_{ijk} A_{ijk}.
\]
This matrix is either real or complex depending on whether we are building the scalar or vector heat equation (\resp).  In particular, we apply a one-step backward Euler approximation to our short-time heat equations (\textsc{Steps I--III} of \algref{vector_heat_method}) to obtain
\[
   \begin{array}{rcl}
      (\Msf - t\LsfNabla)\Ysf &=& \Ysf_0, \\
      (\Msf - t\Lsf)\usf &=& \usf_0, \\
      (\Msf - t\Lsf)\phi &=& \phi_0. \\
   \end{array}
\]
Here the vectors \(\Ysf_0 \in \CC^{|V|}\), \(\usf_0,\phi_0 \in \RR^{|V|}\) describe the source data.  For instance, if the source is a collection of points \(i_1, \ldots, i_k \in V\) with associated vectors \(X_{i_1}, \ldots, X_{i_k}\), then
\[
 \Ysf_0 = \sum_{i=1}^k X_k \delta_k,\qquad
 \usf_0 = \sum_{i=1}^k |X_k| \delta_k,\qquad
 \phi_0 = \sum_{i=1}^k \delta_k,
\]
where \(\delta_k\) is the Kronecker delta at vertex \(k\).  The final result is obtained by evaluating \(\usf_i \Ysf_i/\phi_i |\Ysf|_i\) at each vertex \(i \in V\) (\StepIV{}).

\subsection{Tangent Intrinsic Delaunay}
\label{sec:TangentIntrinsicDelaunay}

\setlength{\columnsep}{-.5em}
\setlength{\intextsep}{0em}
\begin{wrapfigure}{r}{67pt}
   \includegraphics{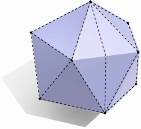}
\end{wrapfigure}
Although this discretization already works quite well, we can improve robustness and accuracy by building the matrices \(\Lsf\), \(\LsfNabla\), and \(\Msf\) with respect to the \emph{intrinsic Delaunay triangulation} of the given mesh~\cite{Bobenko:2007:DLB}.  The basic idea is to construct a different triangulation of the same piecewise Euclidean surface (\ie, without changing the geometry) which leads to better numerical behavior for operators like the Laplacian.  Conceptually, edges may cross multiple faces of the original mesh (see inset), but in practice we need only keep track of the usual connectivity information: which edges are connected to which vertices?  Since the vertex set is preserved, any solution computed on the Delaunay mesh can be directly copied back to the vertices of the original mesh.  However, we will need to augment this construction to keep track of tangent-valued data.

Let \(k,l\) denote the vertices opposite a given edge \(ij\).  The basic intrinsic Delaunay algorithm iteratively flips any edge \(ij\) where the angle sum \smash{\(\theta_k^{ij} + \theta_l^{ji}\)} is greater than \(\pi\), or equivalently, where the cotan weight \(w_{ij}\) is negative.  After a flip, edge \(kl\) is assigned a new length \(\ell_{kl}\), equal to the distance between \(k\) and \(l\) along the previous triangles \(ijk\) and \(jil\).  This length can be computed via the law of cosines---see \citet{fisher2006algorithm} for further details.  We make a small but important modification to this algorithm: after each flip, we also compute the angles \(\varphi\) encoding the outgoing direction of the flipped edge \(kl\) relative to its endpoints.  In particular, we set
\[
   \begin{array}{rcl}
      \varphi_{lk} &\gets& \varphi_{lj} + 2\pi\theta_l^{jk}/\Theta_l, \\
      \varphi_{kl} &\gets& \varphi_{ki} + 2\pi\theta_k^{il}/\Theta_k. \\
   \end{array}
\]
In other words we add the (normalized) angle between the preceding edge and the new edge to the angle of the preceding edge; here the angles \(\smash{\theta_l^{jk}, \theta_k^{il}}\) can be computed from the updated lengths.  This way, we preserve the local polar coordinate systems throughout the flipping process, and therefore know how to map tangent data computed on the intrinsic Delaunay triangulation back to the original mesh: simply copy the polar coordinates \((r_i,\varphi_i)\).

\begin{figure}
   \centering
   \includegraphics[width=\linewidth]{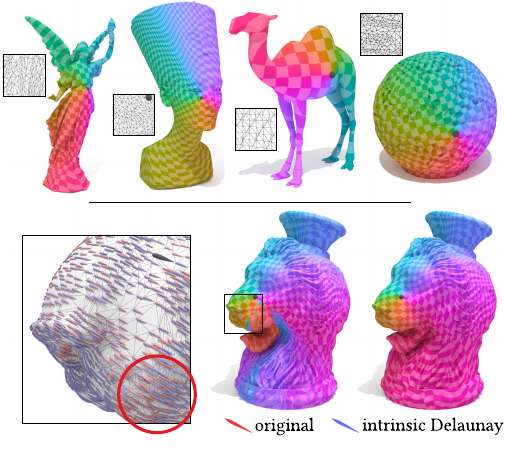}
   \caption{We often obtain high-quality results even on non-Delaunay meshes \figloc{(top)}.  Occasionally, however, transported fields can have improperly oriented vectors \figloc{(inset)}, here causing errors in the log map \figloc{(bottom center)}.  By keeping track of tangent spaces during intrinsic Delaunay flips, we obtain a high-quality solution \figloc{(bottom right)} without having to change the input geometry.\label{fig:intrinsic_delaunay_maps}}
\end{figure}

This procedure is useful not only for our algorithm, but any algorithm that seeks to improve the quality and reliability of tangent vector field processing without altering the geometry of the input mesh.  For instance, the discrete connection Laplacian \(\LsfNabla\) may be indefinite since (unlike the cotan Laplacian \(\Lsf\)) it cannot simply be interpreted as the restriction of the smooth connection Laplacian \(\ConnectionLaplacian\) to the subspace of piecewise linear functions.  The intrinsic Delaunay condition ensures that \(\LsfNabla\) is semidefinite, since the intrinsic cotan weights \(w_{ij}\) are guaranteed to be nonnegative.  It also ensures that no vector \(X_i\) can be ``flipped,'' in the sense that it will always be a positive linear combination of the neighbors \(r_{ji}X_j\) (and hence in their convex cone).  In practice the intrinsic Delaunay condition is not strictly necessary to obtain high-quality results, but helps to guarantee that problems will not occur, even on pathological inputs (\figref{intrinsic_delaunay_maps}).  Formally understanding further properties of the intrinsic connection Laplacian and associated objects (\eg, special intrinsic vector fields) is an enticing question for future work.

\section{Generalizations}
\label{sec:Generalizations}

The vector heat method easily generalizes to other domains and other kinds of vector-valued data (\secref{OtherVectorBundles}), and can easily be implemented on many data structures beyond triangle meshes (\secref{OtherDiscretizations}).

\subsection{Other Vector Bundles}
\label{sec:OtherVectorBundles}

As stated, the algorithm described in \secref{SmoothFormulation} already applies to any \emph{vector bundle}.  Loosely speaking, a vector bundle is a manifold \(M\) with a copy of the same vector space \(V\) at each point.  A choice of vector at each point of \(M\) is called a \emph{section} of the bundle---for instance, a tangent vector field \(X\) is a section of the tangent bundle \(TM\).  As hinted at in \secref{ParallelTransportAndConnections}, the connection \(\nabla\) defines what it means for nearby vectors to be parallel.  In general we may want to change the domain (\ie, the choice of manifold \(M\)), the type of vector data (\ie, the choice of vector space \(V\)), or the notion of what it means for vectors to be parallel (\ie, the choice of connection \(\nabla\)).  These choices ultimately determine the operators \(\Delta\) and \(\ConnectionLaplacian\), which are all we need to formulate \algref{vector_heat_method}.  An elementary example is the \emph{trivial real line bundle}, where the vector space is just \(V = \R\), \ie, sections are just real-valued functions, and parallel transport simply copies values from one point to another.  In this case the vector heat method reduces to the scalar interpolation scheme described in \secref{ScalarInterpolation}---some more interesting examples are given below.

\subsubsection{Differential 1-Forms}
\label{sec:Differential1Forms}

One vector bundle common in applications is the \emph{cotangent bundle} \(T^*\!\Mfld\), whose sections are called \emph{differential 1-forms}.  In this context, it is often easiest to discretize the Hodge Laplacian \(\Delta_1 := d \delta + \delta d\). To obtain a corresponding connection Laplacian \(\ConnectionLaplacianOne\), we will employ the \emph{Weitzenb\"{o}ck identity}, which for a smooth surface \(M\) with Gaussian curvature \(K\) says that
\[
   \ConnectionLaplacianOne \alpha = \Delta_1\alpha + \tfrac{1}{2}K\alpha
\]
for any 1-form \(\alpha\).  We can therefore obtain a discrete connection Laplacian by adding a matrix representing the curvature term to an existing discretization of the Hodge Laplacian.  In particular, let
\[
   \Omega_{ijk} := \pi - (\tilde{\theta}_i^{jk} + \tilde{\theta}_j^{ki} + \tilde{\theta}_k^{ij})
\]
be the total Gaussian curvature of triangle \(ijk\) (see \citet[Section 5.2]{Sharp:2018:VSC}).  Similarly, let
\[
   \Omega_{ij} := \tfrac{1}{3}( \Omega_{ijk} + \Omega_{jil})
\]
be the total curvature in the barycentric diamond around an edge \(ij\) with opposite vertices \(k, l \in V\), which has area
\[
   A_{ij} := \tfrac{1}{3}( A_{ijk} + A_{jil} ).
\]

\setlength{\columnsep}{1em}
\setlength{\intextsep}{0em}
\begin{wrapfigure}{r}{109pt}
  \includegraphics{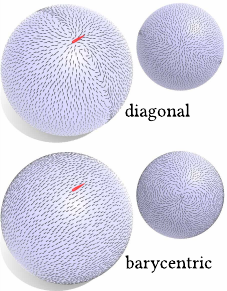}
\end{wrapfigure}
One discrete Hodge Laplacian is provided by \emph{discrete exterior calculus}~\cite{Desbrun:2006:DDF}, defined in terms of simplicial boundary/coboundary operators, in conjunction with diagonal mass matrices \(\star_k\) for differential \(k\)-forms.  One can likewise add a diagonal matrix with entries \(\Omega_{ij}/A_{ij}\) to approximate the connection Laplacian---unfortunately, this approach does not yield the correct result (inset, \figloc{top}); seemingly, the highly local region of support is not sufficient to capture transport.  We therefore replace \(\star_1\) with an approximate Galerkin mass matrix \(\star_1^B\) obtained via one point barycentric quadrature~\cite{Mohamed:2016:CDH}, which has larger support and appears to capture sufficient directional information.  For any two edges $ij$ and $jk$ contained in a common face $ijk \in F$, this matrix has a nonzero entry

\begin{center}
   \vspace{-\baselineskip}
   \begin{tabular}{cc}
      \(\displaystyle (\star_1^B)_{ij,jk} := \displaystyle\frac{|e_{ij}^{\ast}|}{|e_{jk}|}\displaystyle\frac{\cos\left(\xi\right)}{\cos\left(\zeta\right)},\)
      & \text{\raisebox{-.5\height}{\includegraphics{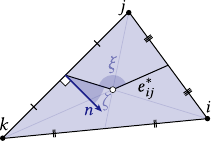}}}
   \end{tabular}
\end{center}

\begin{figure}[b]
   \includegraphics{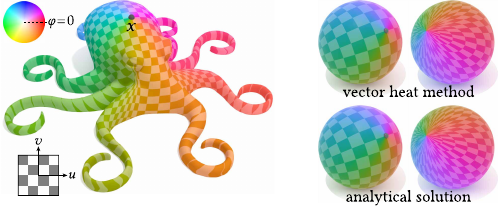}
   \caption{\figloc{Left:} the \emph{logarithmic map} provides a coordinate system on the surface, relative to a chosen origin \(x\). \figloc{Right:} on the sphere we easily compute the correct log map; note that far from the source even the analytical log map can be highly skewed.\label{fig:sphere_comparison}}
\end{figure}

\vspace{.5\baselineskip}\noindent where $\xi$ is the angle between the barycentric dual edges \(e_{ij}^\ast\) and \(e_{jk}^\ast\) (\ie, the vectors from the edge midpoints to the face barycenter), and \(\zeta\) is the angle between the normal \(n\) of edge \(jk\) and the dual edge $e_{jk}^{\ast}$ (see inset figure). To discretize the term \(\tfrac{1}{2}K\) we then build a matrix \(\Ksf \in \RR^{|E| \times |E|}\) with diagonal entries
\[
   \Ksf_{ij,ij} = \tfrac{\Omega_{ij}}{2A_{ij}}(\star_1^B)_{ij,ij}
\]
for each edge \(ij \in E\), and off-diagonal entries
\[
   \Ksf_{ij,jk} = \tfrac{\Omega_{ijk}}{2A_{ijk}}(\star_1^B)_{ij,jk}
\]
for all pairs of edges \(ij,jk \in E\) that share a triangle.  When used in \algref{vector_heat_method}, this discretization appears to yield the correct solution---for instance, the solution on the sphere above closely matches the analytical solution (compare with \figref{sphere_comparison}, \figloc{bottom}).

\subsubsection{Symmetric Direction Fields}
\label{sec:SymmmetricDirectionFields}

\setlength{\columnsep}{1em}
\setlength{\intextsep}{0em}
\begin{wrapfigure}{l}{110pt}
   \includegraphics[width=110pt]{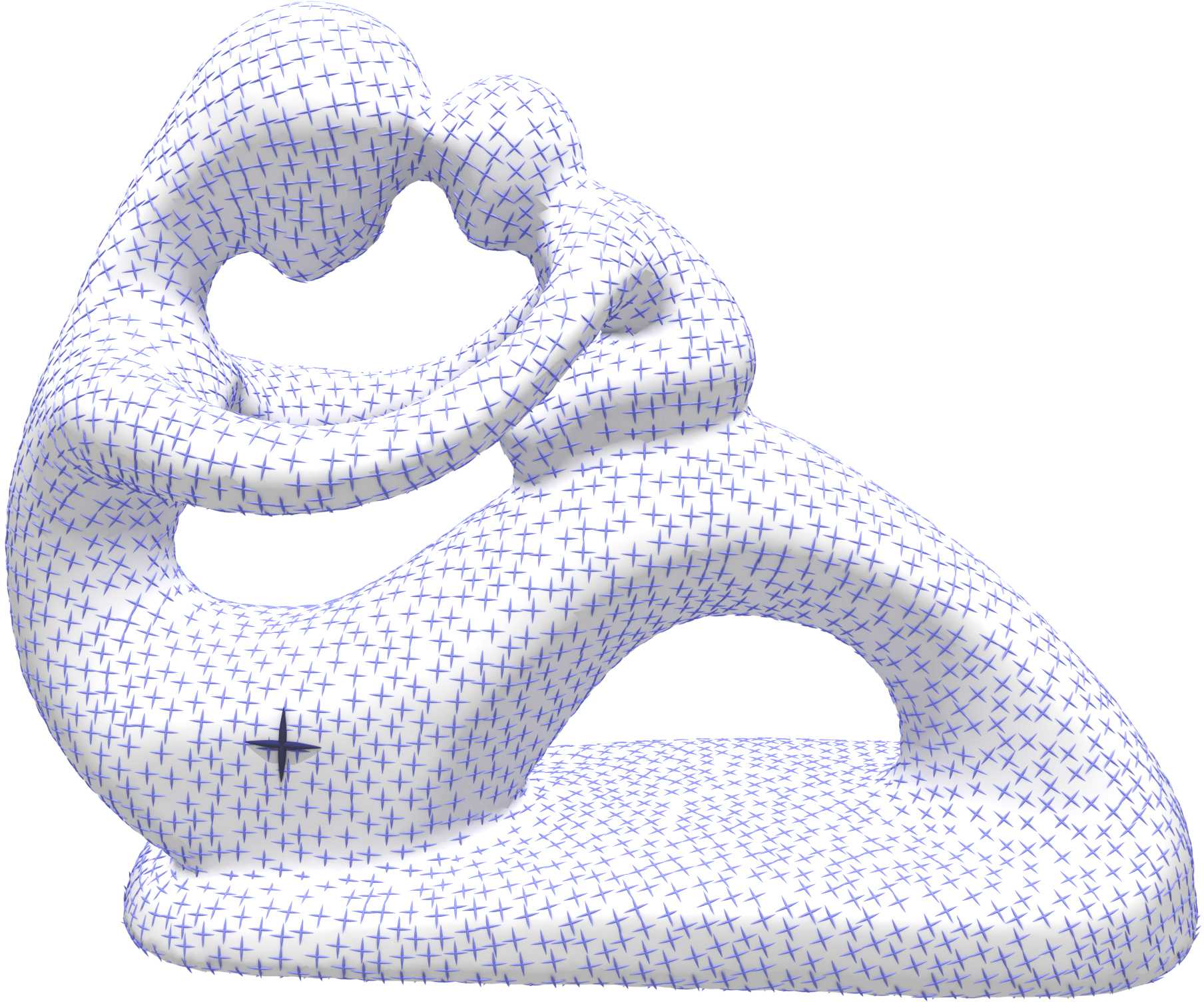}
\end{wrapfigure}
Another important example of vector-valued data in computer graphics and geometry processing are \emph{symmetric direction fields} such as line fields, cross fields, \etc~\cite{Vaxman:2016:DFS}, which play a key role in applications like surface shading~\cite{Knoppel:2015:SPS} and quadrilateral remeshing~\cite{Bommes:2013:QGP}.  Formally, such fields are sections of the \(k\)th tensor power \(TM^{\otimes k}\) of the tangent bundle, where \(k\) determines the degree of symmetry.  In this setting, one can build the connection Laplacian exactly as described in \citet{Knoppel:2013:GOD}---in particular, all one has to do is raise the coefficients \(r_{ij}\) from \eqref{parallel_transport_rotations} to the \(k\)th power, and apply \algref{vector_heat_method} as usual.  The final vectors are given by the \(k\)th complex roots at each vertex, as described in \citet[Sec. 2]{Knoppel:2013:GOD}.  An example is shown in the inset figure.

\subsubsection{Different Connection}
\label{sec:DifferentConnection}

\setlength{\columnsep}{1em}
\setlength{\intextsep}{0em}
\begin{wrapfigure}{r}{97.5pt}
   \includegraphics{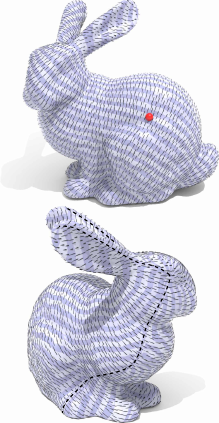}
\end{wrapfigure}
Another possibility is to change the connection \(\nabla\) itself.  In this case, \algref{vector_heat_method} will compute parallel transport along ordinary geodesics, but the notion of what it means to parallel will change.  For instance, setting \(r_{ij} = 1+0\imath\) for all edges \(ij\) simply yields closest point interpolation of complex functions.  A more interesting connection is considered by \citet{Knoppel:2015:SPS}, who compute a global parameterization aligned to a given vector field \(Z\).  Here the Levi-Civita connection \(\nabla\) is replaced by the connection \(\tilde{\nabla} = d - \imath Z^\flat\); in practice this just means that the rotations \(r_{ij}\) are larger for edges that align with \(Z\) (see \citet[Section 3.2]{Knoppel:2015:SPS}).  Using the corresponding connection Laplacian \(\ConnectionLaplacianTilde\) in \algref{vector_heat_method} will yield a \emph{local} field aligned parameterization centered around a given point \(x\), since for any shortest geodesic \(\gamma\) starting at \(x\) the augmented parallel transport map is given by
\[
   \tilde{P}_\gamma = e^{\imath\int_\gamma \langle Z, \gamma^\prime \rangle\ ds},
\]
\ie, a rotation determined by how much the tangent of \(\gamma\) lines up with \(Z\).  Hence, the angle of the transported field \(\eta(y) := \mathrm{arg}(\tilde{P}_{\gamma_{x \to y}} 1)\) will be a scalar function increasing along \(Z\) (see inset).  The gradient of \(\eta\) will therefore be closely aligned with \(Z\) near the source point \(x\); any non-integrability is effectively dealt with by pushing it out toward the cut locus (dashed line), rather than inserting new singularities (as in \citet{Knoppel:2015:SPS}) or globally projecting onto a more integrable field (as in \citet{Ray:2006:PGP}).

\begin{figure*}
   \centering
   \includegraphics[width=\linewidth]{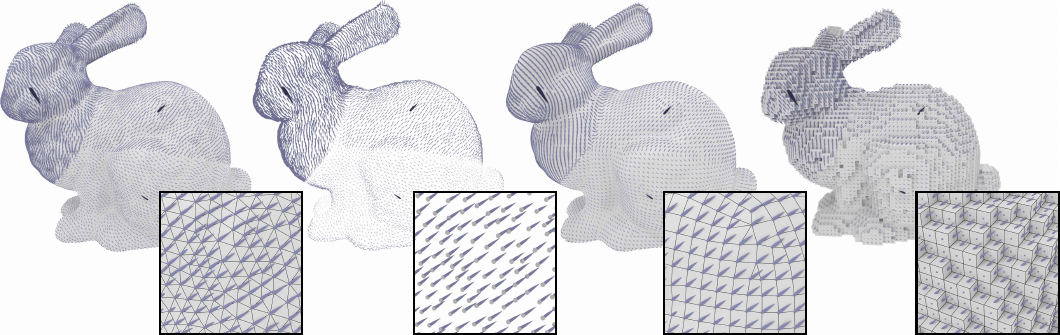}
   \caption{To implement our method on a given geometric data structure, we essentially just need a scalar Laplacian and a notion of tangent spaces at each point or vertex.  Here we show the solution for three sources of different magnitudes on a triangle mesh, point cloud, polygon mesh, and voxelization.\label{fig:various_discretizations}}
\end{figure*}

\subsection{Other Discretizations}
\label{sec:OtherDiscretizations}

There is no fundamental reason why we must use triangle meshes to discretize the vector heat method: any geometric representation that admits a discretization of the scalar Laplace-Beltrami operator \(\Delta\) and the connection Laplacian \(\ConnectionLaplacian\) will suffice.  Discrete Laplacians have been developed for a wide variety of domains, including point clouds~\cite{Liu:2012:PMH}, polygon meshes~\cite{Alexa:2011:DLG}, subdivision surfaces~\cite{deGoes:2016:SEC}, tetrahedral meshes~\cite{Belyaev:2015:VPB}, spline surfaces~\cite{Nguyen:2016:CFE}, and \emph{digital surfaces}, \ie, voxelizations~\cite{Caissard:2017:HKL}, all of which have been used to implement the scalar heat method (see either the preceding references, or \citet{Crane:2013:GHN}).

Given a scalar Laplacian, a connection Laplacian can be obtained by following the same strategy used for triangle meshes (\secref{DiscreteLaplaceOperators}): for any pair of nearby nodes \(i\) and \(j\) (representing vertices, points, \etc), determine the transformation between tangent spaces.  For a surface embedded in \(\mathbb{R}^n\), this transformation is just the smallest rotation between tangent planes, and can be encoded by a unit complex number \(r_{ij}\).  Then simply multiply the off-diagonal entries \(\Lsf_{ij}\) by the values \(r_{ij}\) to obtain the connection Laplacian \(\smash{\Lsf^\nabla}\).  If \(\Lsf\) was symmetric, \(\Lsf_{ij}\) will be Hermitian (assuming \(\smash{r_{ji} = r_{ij}^{-1}}\)), and will hence exhibit the properties discussed in \secref{BasicProperties}).  We consider three specific cases in detail.

\subsubsection{Polygon Meshes}
\label{sec:PolygonMeshes}

For surface meshes comprised of general, possibly non-planar polygons, we augment the discrete Laplacian of \citet{Alexa:2011:DLG} (and use the same mass matrix \(\Msf\)).  In this setting we need a transport coefficient for all pairs of vertices \(i\), \(j\) contained in each polygon---not just those connected by an edge.  We therefore define extrinsic tangent planes, by picking any reasonable normal direction \(N_i\) at each vertex \(i \in V\), and any direction \(E_i\) orthogonal to \(N_i\) that serves as the zero direction.  Letting \(R_{ij} \in \mathrm{SO}(3)\) denote the smallest rotation taking plane \(i\) to plane \(j\), the rotations \(r_{ij}\) are then determined by the angle from \(R_{ij} E_i\) to \(E_j\) in plane \(j\).  These values are used to modify the off-diagonal entries as described above.  \figref{various_discretizations}, \figloc{center right} shows an example on a quad-dominant mesh containing non-planar quads and pentagons, of the type commonly used in numerical simulation.

\subsubsection{Point Clouds}
\label{sec:PointClouds}

As in \citet[Section 3.2.3]{Crane:2013:GHN}, we use the positive semidefinite point cloud Laplacian of \citet{Liu:2012:PMH} to implement our algorithm on unstructured point cloud data.  In this setting we must already estimate tangent planes at each point in order to build the scalar Laplacian; the transport coefficients can therefore be computed exactly as in the polygonal case: pick a direction \(E_i\) at each tangent plane and compute the rotations \(r_{ij}\) from \(R_{ij}E_i\) to \(E_j\).  As in the scalar case, the mass matrix \(\Msf\) is given by the Voronoi areas associated with points.  An example is shown in \figref{various_discretizations}, \figloc{center left}.

\subsubsection{Voxelizations}
\label{sec:Voxelizations}

Finally, for a voxelized or \emph{digital surface} a discrete Laplacian was recently developed by \citet{Caissard:2017:HKL}.  Here, values are associated with faces (\ie, quads) on the voxelization boundary; the basic principle is the same as in the point cloud case, but normals and areas are carefully estimated based on the voxelization geometry.  (If the voxelization arises from an Eulerian signed distance function, these normals might also be used.)  An example is shown in \figref{various_discretizations}, \figloc{far right}.

\section{Discussion and Validation}
\label{sec:DiscussionAndValidation}

Here we study numerical properties of our method, and compare with other candidate approaches.  Importantly, the main task we consider (parallel transport along shortest geodesics to a given set) is not one previously considered in geometry processing, and hence there are no standard algorithms.  The closest analogy, perhaps, is a recent numerical integrator for closed-form Riemannian metrics rather than discrete meshes~\cite{Louis:2017:FSP}.  For surfaces of revolution we use the exact solution (computed via Clairaut's relation) as a basis for comparison; for more complicated models we compute exact polyhedral geodesics (\ala{} \citet{Surazhsky:2005:FEA}), and apply parallel transport via unfolding, as described by \citet{Polthier:1998:SGP}.  Note that even the polyhedral approach does not yield the true (smooth) solution on coarse meshes or near the cut locus; on fine meshes and away from the cut locus it nonetheless provides a useful benchmark for comparison.  Overall we find that the vector heat method provides an excellent performance/quality trade off, making it well-suited for practical geometry processing tasks; application-specific comparisons are explored in \secref{Applications}.

\begin{figure*}
   \centering
   \vspace{-\baselineskip}
   \includegraphics[width=\linewidth]{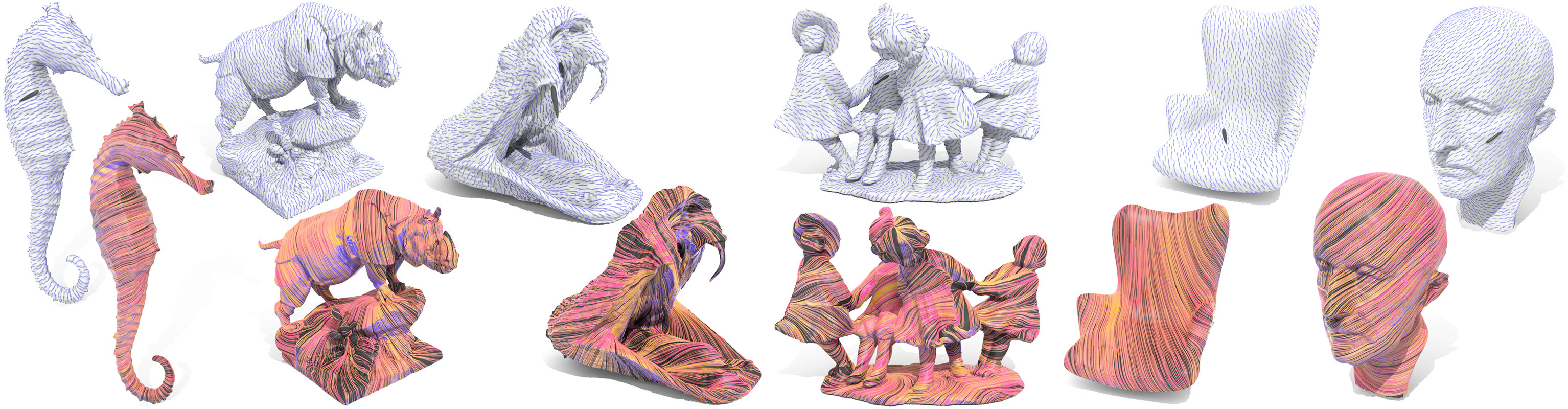}
   \caption{Results of the vector heat method on a variety of models; source is marked by a large arrow. For visualization, vectors are sampled \ala{} \citet{Bridson:2007:FPD}.\label{fig:transport_gallery}}
\end{figure*}

\begin{table}
\begin{tabularx}{\linewidth}{X|XXr}
\textsc{Model} & \textsc{Triangles} & \textsc{Precompute} & \textsc{Solve} \\
\hline
Chair          &  11k               & 0.06\text{ s}       & 0.002\text{ s} \\
Bust           & 100k               & 0.73\text{ s}       & 0.031\text{ s} \\
Children       & 100k               & 0.74\text{ s}       & 0.027\text{ s} \\
Seahorse       & 145k               & 1.25\text{ s}       & 0.047\text{ s} \\
Snake          & 293k               & 2.74\text{ s}       & 0.101\text{ s} \\
Rhino          & 310k               & 4.02\text{ s}       & 0.105\text{ s} \\
\end{tabularx}
   \caption{Timings for Figure \ref{fig:transport_gallery}, including intrinsic Delaunay preprocessing.
   \label{tab:gallery_stats}}
\end{table}

\begin{figure}[b]
   \centering
   \includegraphics{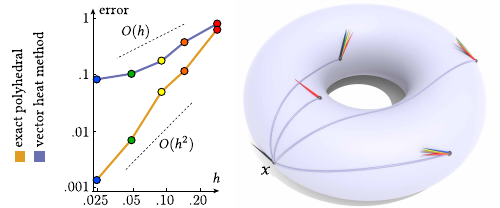}
   \caption{The vector heat method appears to converge linearly.  \figloc{Left:} On progressively finer meshes (ranging from 1250--11250 triangles), we transport a vector from a source \(x\) to several other points and measure the error relative to the true solution on the smooth surface. \figloc{Right:} visualization of vectors transported via the vector heat method; black is reference solution.\label{fig:pointwise_convergence}}
\end{figure}

\subsection{Basic Properties}
\label{sec:BasicProperties}

Which properties of smooth parallel transport are preserved by our discrete algorithm?  For a single point source, one can easily argue that we exactly preserve elementary properties such as linearity (\(P_\gamma(aX + Y) = aP_\gamma X + P_\gamma Y\)), conservation of magnitude (\(|P_\gamma X| = |X|\)), and covariance with respect to rotation, \ie, rotating the initial vector is equivalent to rotating the final solution by the same angle.  A more interesting property is symmetry: in the smooth setting, \(P_{\gamma_{y \to x}} \circ P_{\gamma_{x \to y}} = \mathrm{id}\), \ie, transporting from \(x\) to \(y\) and back again should yield the original vector.  This property turns out to be true in the discrete case as well: since the matrix \(\LsfNabla \in \CC^{|V| \times |V|}\) encoding the connection Laplacian is Hermitian, the solution operator
\[
   \Asf := (\Msf + t \LsfNabla)^{-1}
\]
is also Hermitian.  Letting \(\delta_i\) denote a Kronecker delta at the source vertex \(i \in V\), we can write the parallel vector field corresponding to the vector \(\bsf := z\delta_i\) (for \(z \in \CC\)) as \(\Xsf = \Asf \bsf.\)
The transported vector at any vertex \(j \in V\) can be written as \(\Xsf_j = (\delta_j^T \Xsf)\delta_j\); when we transport this vector back to \(i\), we therefore get
\[
   \delta_i^T (\Asf \Xsf_j) = \delta_i^T \Asf \delta_j (\delta_j^T \Xsf) = z (\delta_i^T \Asf \delta_j)(\delta_j^T \Asf \delta_i) = z \Asf_{ij} \Asf_{ji}.
\]
Since \(\Asf\) is Hermitian, \(\Asf_{ij} = \overline{\Asf}_{ji}\), hence \(\Asf_{ij} \Asf_{ji}\) is real, \ie, the initial vector \(z\) experiences a scaling and no rotation.  But since the overall process preserves scale, the final vector is the same as the initial one.  Further properties of parallel transport (such as equivalence between curvature and monodromy around closed loops) may not hold exactly; likewise, these properties may not hold exactly in the case of multiple sources or curves, since for \(t > 0\) vectors pointing in different directions may result in cancellation of magnitude.  In general, we expect that many properties will at least be preserved under refinement---see \secref{ConvergenceAndAccuracy} for further discussion.

\subsection{Implementation and Performance}
\label{sec:ImplementationAndPerformance}

We implemented our method in C++ using double precision; all timings were taken on a single thread of an Intel Core i7 3.5GHz CPU.  To solve linear systems we prefactored matrices via CHOLMOD~\cite{Chen:2008:ACS} or UMFPACK~\cite{Davis:2004:AUV} and applied backsubstitution for each subsequent problem.  For the basic algorithm (\algref{vector_heat_method}) we need to pre-factor two \(|V| \times |V|\) Laplace matrices (one real, one complex); for each new source set we need only three backsolves, and trivial per-vertex multiplication/division operations.  The (optional) intrinsic Delaunay mesh can be constructed as a preprocess; in practice we find the cost is about the same as one matrix factorization.  On a mesh of $100$k triangles, preprocessing takes about 1 second overall; computing parallel transport from any subsequent source to all points on the surface then takes about $30$ms.  We did not carefully optimize our code, though further accelerations are relatively straightforward: for instance, since both matrices have the same sparsity pattern one could re-use the symbolic factorization; moreover, since there is no dependence among the linear systems, backsubstitution could be applied in parallel.  (See also discussion of fast preconditioners in \secref{RelatedWork}.) In contrast, computing exact polyhedral geodesics and applying parallel transport via unfolding takes about 40x longer on the same mesh (using Kirsanov's implementation of \cite{Surazhsky:2005:FEA}).  One could significantly improve performance of the polyhedral strategy via any number of recent acceleration schemes (such as \cite{Ying:2013:SVG}), or, at the cost of accuracy, by extracting geodesics from a cheaper piecewise linear distance function.  However, even just \emph{tracing} the geodesics from the source point to each vertex (whether using the polyhedral scheme or fast marching) already takes about 10--20x longer than executing our entire method; in general, it would seem quite difficult to develop a polyhedral strategy that is competitive with the diffusion-based approach.

\begin{figure}[b]
   \centering
   \includegraphics{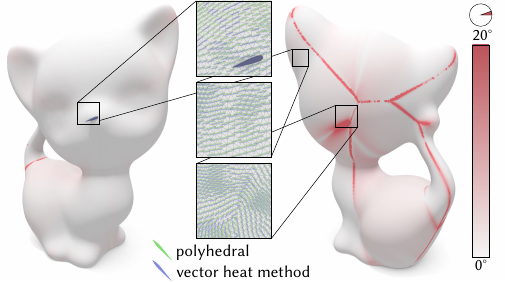}
   \caption{Our algorithm yields very similar results to the brute-force approach of explicitly unfolding triangles along exact polyhedral geodesics.  Away from the cut locus, the difference is typically just a few degrees (\figloc{left, right}).\label{fig:exact_polyhedral_comparison}}
\end{figure}

\begin{figure*}
   \centering
   \includegraphics[width=\textwidth]{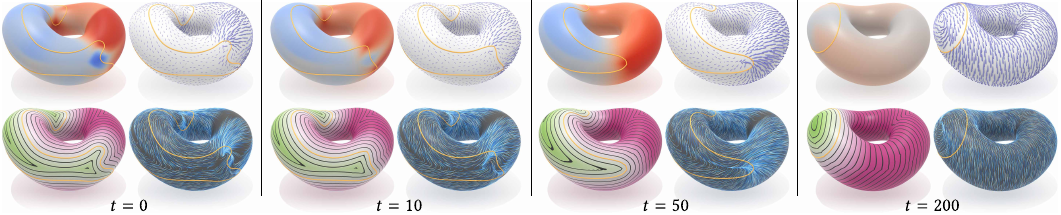}
   \caption{Our method enables fast velocity extension for level set methods. Here we show a sequence of frames from a simple curve-shortening flow, plus a constant tangential term. In each frame, our scalar interpolation scheme provides a closest-point interpolation of normal velocity (\figloc{top left}), resulting in excellent preservation of the signed distance function over long integration times (\figloc{bottom left}); note that we never apply explicit redistancing. Since \algref{vector_heat_method} provides accurate extrapolation of vector data over the entire domain (\figloc{top right}), we can track particles even very far from the interface (\figloc{bottom right}).\label{fig:levelset_flow}}
\end{figure*}

\subsection{Convergence and Accuracy}
\label{sec:ConvergenceAndAccuracy}

The accuracy of any method for computing parallel transport will depend on the resolution and quality of the surface tessellation.  For the vector heat method, we find that using an intrinsic Delaunay triangulation improves quality, and hence apply this technique throughout our examples.  Exact polyhedral schemes also re-tessellate the input by slicing it up into polygonal ``windows'' relative to the given source.  This point of view helps to explain the relative trade offs of the two approaches: the vector heat method re-tessellates the domain at most once (as an optional pre-process), whereas polyhedral schemes must re-tessellate for each new source point.  The vector heat method hence has far better amortized performance, whereas window-based schemes can provide greater accuracy since the domain is effectively meshed along characteristics of the equation being solved (\ie, along geodesics).  Empirically, we observe a convergence rate of roughly \(O(h)\) and \(O(h^2)\) for the two methods, \resp{}, relative to the mean edge length \(h\) (\figref{pointwise_convergence}).  Fields computed via these two methods mainly differ near the cut locus, where \emph{neither} approach can guarantee accurate results---in fact, it is well-known that the exact polyhedral cut locus is a poor approximation of the smooth one~\cite{Itoh:2004:TTA}.  On the whole, accuracy and rates of convergence are in line with the scalar heat method---for a more in-depth analysis, see \citet[Section 4.2]{Crane:2013:GHN}.

\paragraph{Choice of \(t\).} The accuracy of the vector heat method will be affected by the choice of the parameter \(t\).  Here we observe the exact same behavior as for the scalar heat method: if \(h\) is the average spacing between nodes (\eg, the mean edge length in a triangle mesh) then setting
\[
   t = h^2
\]
for both scalar and vector heat flow tends to yield the best accuracy (see \figref{ChoiceOfT} and \cite[Section 3.2.4]{Crane:2013:GHN}).  We use this value throughout all of our examples.

\section{Applications}
\label{sec:Applications}

Fast parallel transport along shortest paths provides a basic foundation on top of which many algorithms can easily be built---here we consider several important examples from geometry processing and simulation.  Implementation of these algorithms via the vector heat method is often much simpler than existing alternatives: mainly just setting up and solving linear systems.  In each case one enjoys a common set of benefits, such as low amortized cost (due to prefactorization) and the ability to generalize to many different geometric data structures (point clouds, polygon meshes, \etc).  To keep discussion concrete, we will describe algorithms primarily in terms of triangulated surfaces.

\subsection{Velocity Extrapolation}
\label{sec:VelocityExtrapolation}

Perhaps the most straightforward application of our method is extrapolation of scalar or vector velocity in the context of free boundary problems; beyond physical simulation, such methods are increasingly used for tasks ranging from shape optimization to semantic shape analysis.  If a signed distance function needs to be updated, one can simply extrapolate the scalar velocity in the normal direction, using the approach described in \secref{ScalarInterpolation}.  To advect auxiliary quantities (color, temperature, particles, \etc), one also needs to extrapolate the tangential velocity, which can be done using \algref{vector_heat_method}.  \figref{levelset_flow} illustrates several features of our extrapolation strategy: for instance, since we solve a global problem we get a well-behaved velocity field far from the interface; in fact, the closest point property ensures that a signed distance function will be nearly preserved even over long integration times.  Note that due to the use of a finite value \(t\), data directly on the interface may not be exactly preserved, but will generally be very close.  In the scalar case, the performance comparison with fast marching is identical to the comparison found in~\cite[Section 4.1]{Crane:2013:GHN}: the cost of our method is dominated by two backsolves, whereas fast marching executes a Dijkstra-like traversal.  Note that one does \emph{not} need to refactor matrices for a changing boundary.  In the vector case, there is not a clear comparison: vector extrapolation is well-studied for Euclidean domains (\eg, \cite{Adalsteinsson:1999:FCE}), but these methods do not immediately generalize to curved surfaces due to the added complexity of parallel transport (especially on data structures like point clouds).

\subsection{Logarithmic Map}
\label{sec:LogarithmicMap}

\begin{figure}
  \includegraphics[width=\columnwidth]{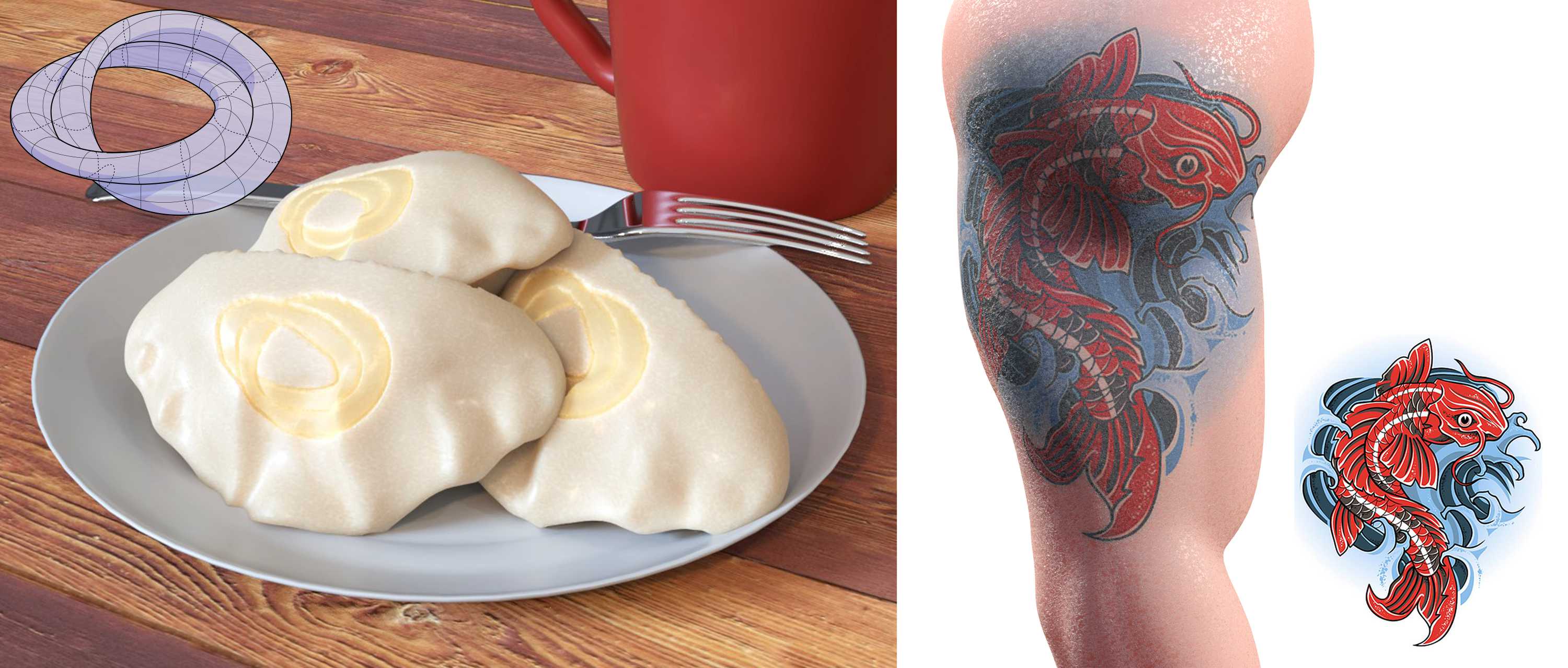}
   \caption{A noise-free log map allows us to place delicious decals \figloc{(left)} or tantalizing tattoos \figloc{(right)} on a surface without having to worry about difficult issues like where to place cuts.\label{fig:decal_examples}}
\end{figure}

At any point \(x\) of a closed surface \(M\), the \emph{exponential map} \(\Exp_x(r v)\) yields the point \(y \in M\) obtained by walking in the unit tangent direction \(v\) and continuing along a geodesic for a distance \(r\).  The \emph{logarithmic map} \(\log_x\) is the inverse operation: given any point \(y\) (away from the cut locus), it finds the smallest distance \(r\) and corresponding unit vector \(v\) such that \(\mathrm{exp}_x(rv) = y\), in analogy with the ordinary logarithm and exponential.  If we encode \(v\) as an angle \(\varphi\), then we essentially have polar coordinates \((r,\varphi)\) relative to an origin \(x\).  As observed by \citet{Schmidt:2006:IDC}, the logarithmic map (referred to there as the exponential map) is useful for a large number of tasks in geometry processing, such as interactive shape editing~\cite{Schmidt:2010:MIR} and texture decaling (\figref{decal_examples}).  A \emph{global} logarithmic map also helps translate algorithms from Euclidean space to curved domains---for instance, in \secref{KarcherMeans} the log map enables us to easily compute generalized centers of mass, by identifying points on the surface with vectors in the tangent plane.

\begin{figure}[b]
   \centering
   \vspace{-2\baselineskip}
   \includegraphics[width=\columnwidth]{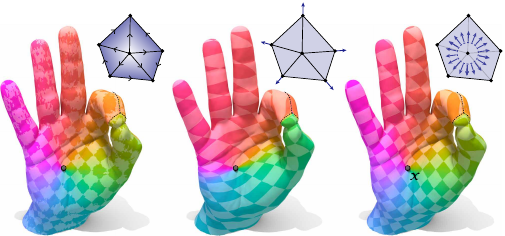}
   \caption{To get a high-quality log map (here, from a source point \(x\)) one must carefully discretize the distance gradient. \figloc{Left:} simply taking the gradient of a given distance function yields numerical noise. \figloc{Middle:} naive parallel transport of vectors emanating from the source yields global anisotropy. \figloc{Right:} proper discretization of initial conditions yields a smooth and accurate map, where the only remaining noise is near the cut locus (dashed line).\label{fig:gradient_discretization_comparison}}
\end{figure}

\setlength{\columnsep}{1em}
\setlength{\intextsep}{0em}
\begin{wrapfigure}{r}{65pt}
  \includegraphics{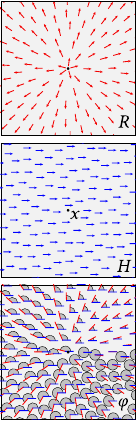}
\end{wrapfigure}
How can we compute the log map on a surface?  For polar coordinates \((r,\varphi)\) in the Euclidean plane, \(\varphi\) can be expressed as the angle between a horizontal direction \(H\) and a radial vector field \(R\) emanating from the origin.  Likewise, on a curved surface the radial vector field \(R\) is given by the gradient of the geodesic distance to a source point \(x\), and the ``horizontal'' vector field \(H\) is obtained by transporting any unit vector at \(x\) to every other point.  The angular coordinate of the log map is then the angle from \(H\) to \(R\); the radial component is just the geodesic distance.  We compute the horizontal field by applying \algref{vector_heat_method} as usual, where the choice of initial vector determines the zero direction (\(\varphi=0\)).  Obtaining the radial field \(R\) is more challenging: if we simply take derivatives of a per-vertex distance function, we get numerical noise (\figref{gradient_discretization_comparison}, \figloc{left}).  Explicitly smoothing this field is not an attractive option, since it will distort features like the cut locus, and generally degrade the accuracy of subsequent computations (\eg, when computing Karcher means).

Instead, we can apply our parallel transport algorithm (\algref{vector_heat_method}) to a small circle of outward pointing normals around the source vertex \(i\).  Here, care must be taken in formulating the initial conditions \(R^0\) for the vector diffusion step: simply setting \(\smash{R^0_j}\) to the outward pointing direction \(-e_{ji}/|e_{ji}|\) at each neighbor \(j\) can lead to anisotropy in the resulting map (\figref{gradient_discretization_comparison}, \figloc{middle}).  We instead derive initial conditions by carefully projecting unit normals on a circle of radius \(\veps\) around the source vertex onto piecewise linear basis functions (as discussed in \appref{DistanceGradientDiscretization}).  This approach yields a log map which is both accurate and smooth (\figref{gradient_discretization_comparison}, \figloc{right}).  Note that since both \(H\) and \(R\) are unit vector fields we do not need to interpolate magnitudes.  The radial coordinate \(r\) (corresponding to the geodesic distance) can be computed using any method; we simply integrate \(R\) by solving the Poisson equation \(\Delta r = \nabla \cdot R\), requiring only one additional prefactorization and backsubstitution.  In particular, to evaluate the right hand side of this equation, we first map vectors \(R_i\) at vertices to integrated values \(R_{ij} \in \RR\) per edge, by averaging the inner product with the edge vector:
\[
      R_{ij} := \tfrac{1}{2}( \langle e_{ij}, X_i \rangle + \langle -e_{ji}, X_j \rangle )
\]
(see \citet[Section 3.2]{Knoppel:2015:SPS} for further discussion).  Keeping in mind that \(R_{ij} = -R_{ji}\), the total divergence \(\nabla \cdot R\) for vertex \(i\) can then be expressed as
\[
   (\nabla \cdot R)_i = \sum_{ij \in E} w_{ij} R_{ij},
\]
where \(w_{ij}\) are the cotan weights from \secref{DiscreteLaplaceOperators}.  The final log map is encoded via Cartesian coordinates \((u,v) := r_i(\cos\varphi_i,\sin\varphi_i)\) at each vertex \(i \in V\).

\begin{figure}
   \centering
   \includegraphics[width=\columnwidth]{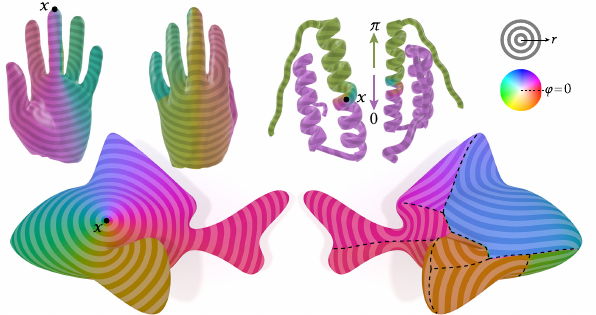}
   \caption{Our log map is globally accurate---even on long skinny models where most points are reached by traveling in nearly identical directions \figloc{(far right)}.  On a high-resolution mesh we nicely resolve the cut locus (dashed line).\label{fig:schmidt_comparison_set}}
\end{figure}

\begin{figure}
   \centering
   \includegraphics[width=\columnwidth]{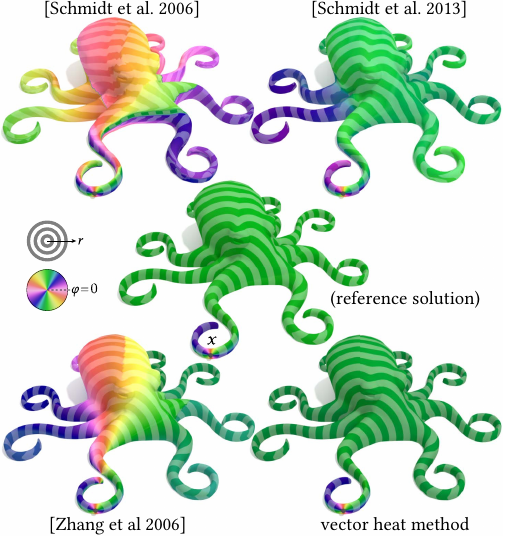}
   \caption{Since the vector heat method is based on a straightforward discretization of a smooth formulation, it leads to a close approximation of the true log map.  Previous approximations exhibit not only small local errors, but large global inaccuracies---especially in the angular coordinate \(\varphi\).  (Reference solution is computed via the exact polyhedral scheme.)\label{fig:octopus_salad}}
\end{figure}

Previous approaches aim to compute a log map only in a neighborhood around the source using \emph{extrinsic} approximations ~\cite{Schmidt:2006:IDC,brun2007manifolds,melvaer2012geodesic}, making them inaccurate over longer distances, and precluding isometry invariance.  For instance, \citet{Zhang:2006:VFD} simply project the closest 25\% of vertices onto the local tangent plane which can fail badly for highly curved surfaces; Dijkstra-like algorithms can also deviate wildly on highly curved domains (see \figref{octopus_salad} for comparisons).  As a result, such approximations cannot reliably be used for algorithms that require global information, such as computation of Karcher means (\secref{KarcherMeans}) or intrinsic landmarks (\secref{OrderedIntrinsicLandmarks}).  Our method nicely resolves the map over the whole surface, all the way up to the cut locus (\figref{schmidt_comparison_set}).  It is also quite competitive in terms of performance since it simply needs to execute highly optimized backsubstitution operations, rather than a Dijkstra-like traversal (see discussion in \secref{RelatedWork}).  When a parameterization is desired only in a small region (\eg, when computing centroidal Voronoi diagrams), further speedups might be achieved by applying the localized Cholesky strategy of~\citet{herholz2017localized}, which fits perfectly into our framework.  \figref{decal_examples} shows some simple examples where our log map is used to add texture decals to a surface.  Note that in the smooth setting the log map is only well-defined for closed surfaces---nonetheless, our method still works nicely on surfaces with boundary, especially within the image of the exponential map.

\begin{algorithm}
\caption{Logarithmic Map}
   \textbf{Input:} A point \(x \in M\) and zero direction \(H_0 \in T_x M\). \\
   \textbf{Output:} A function \((u,v): M \to \RR^2\). \\
\begin{enumerate}[I.]
   \item Transport \(H_0\) to all other points to get \(H\), via \algref{vector_heat_method}.
   \item Compute \(R\) using \algref{vector_heat_method}, with initial conditions from \appref{DistanceGradientDiscretization}.
   \item Compute the angle \(\varphi\) from \(H\) to \(R\) at each point.
   \item Solve the Poisson equation \(\Delta r = \nabla \cdot R\).
   \item Evaluate \((u,v) := r(\cos\varphi,\sin\varphi)\) at each point.
\end{enumerate}
\label{alg:log_map}
\end{algorithm}

\setlength{\columnsep}{1em}
\setlength{\intextsep}{0em}
\begin{wrapfigure}{r}{94pt}
  \includegraphics{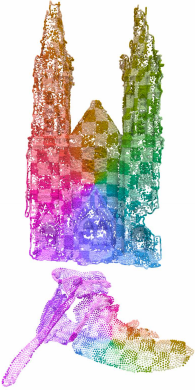}
\end{wrapfigure}
\paragraph{Point cloud log map.} As an example of how the diffusion-based approach easily extends to other geometric data structures, we implemented the log map directly on unstructured point clouds.  Here, computation of the horizontal vector field \(H\) is straightforward: just apply \algref{vector_heat_method} to a unit vector at the source point \(x\); the radial field \(R\) can be computed using the same initial conditions described by \citet{Lin:2014:GDF}: transport the vector from \(x\) to the tangent space at each neighbor \(y\) (via an extrinsic rotation) and project onto the tangent space.  The angular component is then given by the angle from \(H\) to \(R\); the radial component is obtained by solving the Poisson equation \(\Delta r = \nabla \cdot R\) exactly as described in \citet[Section 3.2.3]{Crane:2013:GHN}.  Two examples on scanned data are shown in the inset.

\begin{figure*}
   \centering
   \includegraphics[width=\textwidth]{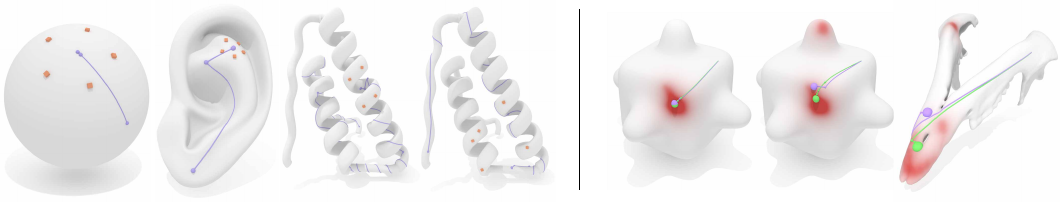}
   \caption{Our globally accurate logarithmic map can be used to compute centers of data on a curved surface. \figloc{Left:} given a collection of points, we iteratively compute the Karcher mean; on simple geometries we reproduce the expected solution in just 2-3 iterations, and the algorithm generalizes to more complex geometries while still needing $< 20$ iterations. \figloc{Right:} We can also efficiently compute the centers of distributions; here we show both the Karcher mean (purple) and the geometric median (green). The Karcher mean is significantly influenced by outliers, while the geometric median is not.\label{fig:karcher_mean_combined}}
\end{figure*}

\subsection{Karcher Means and Geometric Medians}
\label{sec:KarcherMeans}

\begin{algorithm}[b!]
\caption{Karcher Mean}
   \textbf{Input:} A collection of points \(y_1, \ldots, y_n \in M\). \\
   \textbf{Output:} A point \(m \in M\). \\
\begin{enumerate}[I.]
   \item Pick a random initial guess \(m^0 \in M\).
   \item Until the vector \(v\) has sufficiently small norm:
      \begin{enumerate}
         \item Compute the log map at \(m^k\) via \algref{log_map}.
         \item Evaluate the update vector \(v = \tfrac{1}{n}\sum_i \log_{m^k}(y_i)\).
         \item Compute \(m^{k+1} = \Exp_{m^k}(\tau v)\), \ie, walk forward along \(v\) for time \(\tau\).
      \end{enumerate}
\end{enumerate}
\label{alg:karcher_mean}
\end{algorithm}

Given a set of points $y_1, \ldots, y_n \in M$, any minimizer of the energy
\begin{equation}
   \label{eq:karcher_energy}
   E(x) := \frac{1}{2n} \sum_{i=1}^n d(x, y_i)^p
\end{equation}

\setlength{\columnsep}{1em}
\setlength{\intextsep}{0em}
\begin{wrapfigure}{r}{120pt}
   \includegraphics{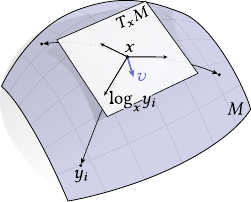}
   \caption{At any point \(x \in M\), the gradient \(v\) of the Karcher mean energy is just the sum of the logarithms of all the points \(y_i\).}
\end{wrapfigure}
\noindent provides a notion of center.  For \(p=2\), such minimizers are called \emph{Karcher means}; in Euclidean space, just the centroid or arithmetic mean. (If the Karcher mean is unique, it is known as the \emph{Fr\'{e}chet mean}.) For \(p=1\), minimizers are known as \emph{geometric medians}, and tend to be more robust to outliers.

Though algorithms have been developed for finding Karcher means on special geometries \cite{buss2001spherical} or other notions of weighted averages on surfaces \cite{panozzo2013weighted}, to date there has been no practical algorithm for accurately computing Karcher means on general surfaces.  Likewise, the geometric median has been considered in the space of images~\cite{fletcher2009geometric}, but no efficient algorithms are known on discrete geometric domains (meshes, point clouds, \etc).  Our algorithm for computing the log map (\secref{LogarithmicMap}) enables a straightforward and efficient strategy for minimizing the energy \(E\).  In the case of the Karcher mean (\(p=2\)), we just iteratively evaluate the update vector
\begin{equation}
   v \gets \frac{1}{n} \sum_{i=1}^n \Log_{m^k}(y_i), \qquad m^{k+1} \gets \Exp_{m^k}(\tau v),
\label{eq:karcher_iteration}
\end{equation}
\noindent where $\tau > 0$ is a step size.  (For instance, if the domain is \(\RR^N\), this algorithm immediately converges to the centroid for \(\tau = 1\).) For the geometric median (\(p=1\)) we simply need to replace the expression for \(v\) with a convex combination \(\sum_i \omega_i \Log_{m^k}(y_i)/\sum_i \omega_i\), where the coefficients \(\omega_i := 1/d(m^k,y_i)\) can also be computed from the log map---this strategy is known as the \emph{Weiszfeld algorithm}~\cite{weiszfeld1937point}.  The log map is computed once per iteration, via the algorithm described in \secref{LogarithmicMap}.  To evaluate the exponential map, we simply trace a geodesic at \(m^k\) along the surface in the direction \(v\) for time \(\tau\) (\ala{} \citet{Polthier:1998:SGP}, in the case of triangle meshes).  In practice this scheme tends to converge in no more than $20$ iterations (and far fewer on simple models); in all our examples, the initial guess \(m^0 \in M\) is chosen completely at random.  The cost of each iteration is dominated by the backsolves to compute the log map.  Line search can slightly reduce the number of steps, at the cost of additional solves; for most examples we use \(\tau = 1\) and no line search.

\begin{figure*}
   \centering
   \includegraphics{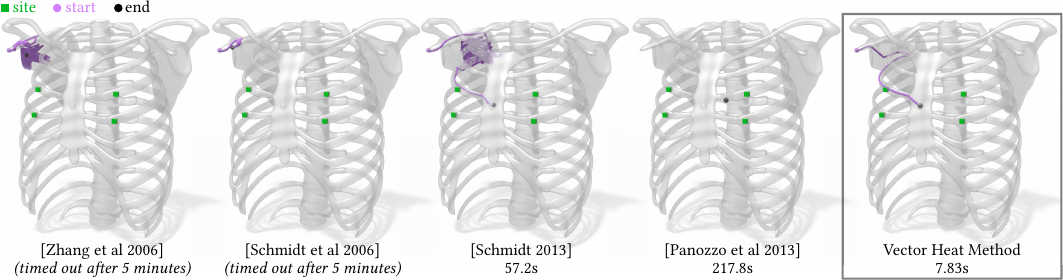}
   \caption{Here we compute the Karcher mean of the four green points, which should appear at the center of the sternum.  Previous local approximations of the log map yield poor behavior, with iterates failing to converge \figloc{(far left, center left)} or wandering around randomly until line search pushes the solution toward a local minimum \figloc{(center)}.  Other algorithms compute only approximate averages, which may not respect symmetries of the problem \figloc{(center right)}.  The global accuracy of the log map provided by the heat method guides the algorithm to the correct solution in just a few iterations \figloc{(far right)}.\label{fig:KarcherMeanComparison}}
\end{figure*}

In fact, since we know the log map over the entire surface, the number of points \(p_i\) has a negligible effect on the cost of computation (just taking a weighted sum).  We can therefore apply the same method directly to arbitrary distributions \(\rho: \M \to \RR_{>0}\) (as depicted in \figref{karcher_mean_combined}, \figloc{right}); in this case, we can define a center
\begin{equation}
  m(\rho) = \argmin_x \int_\M \rho(y)d(x, y)^p~ dy.
\end{equation}
The only change to the algorithm is that we now take a weighted average over \emph{all} vertices, using weights \(\Msf_{ii} \rho_i\), where \(\rho_i \in \RR_{>0}\) is the density at each vertex, and \(\Msf\) is the mass matrix.

The globally accurate approximation of the log map provided by the vector heat method allows us to reliably obtain the correct result (\figref{KarcherMeanComparison}, \figloc{far right}).  Although previous methods for \emph{locally} approximating the log map can also be used to implement this algorithm~\cite{Zhang:2006:VFD,Schmidt:2006:IDC,Schmidt:2013:SP}, they are not accurate enough \emph{globally} to produce the desired behavior: iterates either wander around randomly and fail to converge, or converge only because line search eventually steers an inaccurate guess toward a local minimum (\figref{KarcherMeanComparison}).  The method of \citet{panozzo2013weighted} takes a different, non-iterative approach to computing averages on surfaces, but does not produce the true Karcher mean, and may not even respect basic symmetries (\figref{KarcherMeanComparison}, \figloc{center right}).  Likewise, simply using a smooth, somewhat parallel vector field to construct the log map also does not work, as shown in \figref{smoothest_field_comparison}, \figloc{right}.  To date, the vector heat method appears to be the only way to compute Karcher means on surfaces.
\subsection{Geodesic Centroidal Voronoi Diagrams}
\label{sec:GeodesicCentroidalVoronoiDiagrams}

\setlength{\columnsep}{1em}
\setlength{\intextsep}{0em}
\begin{wrapfigure}{r}{125pt}
   \centering
   \includegraphics[width=125pt]{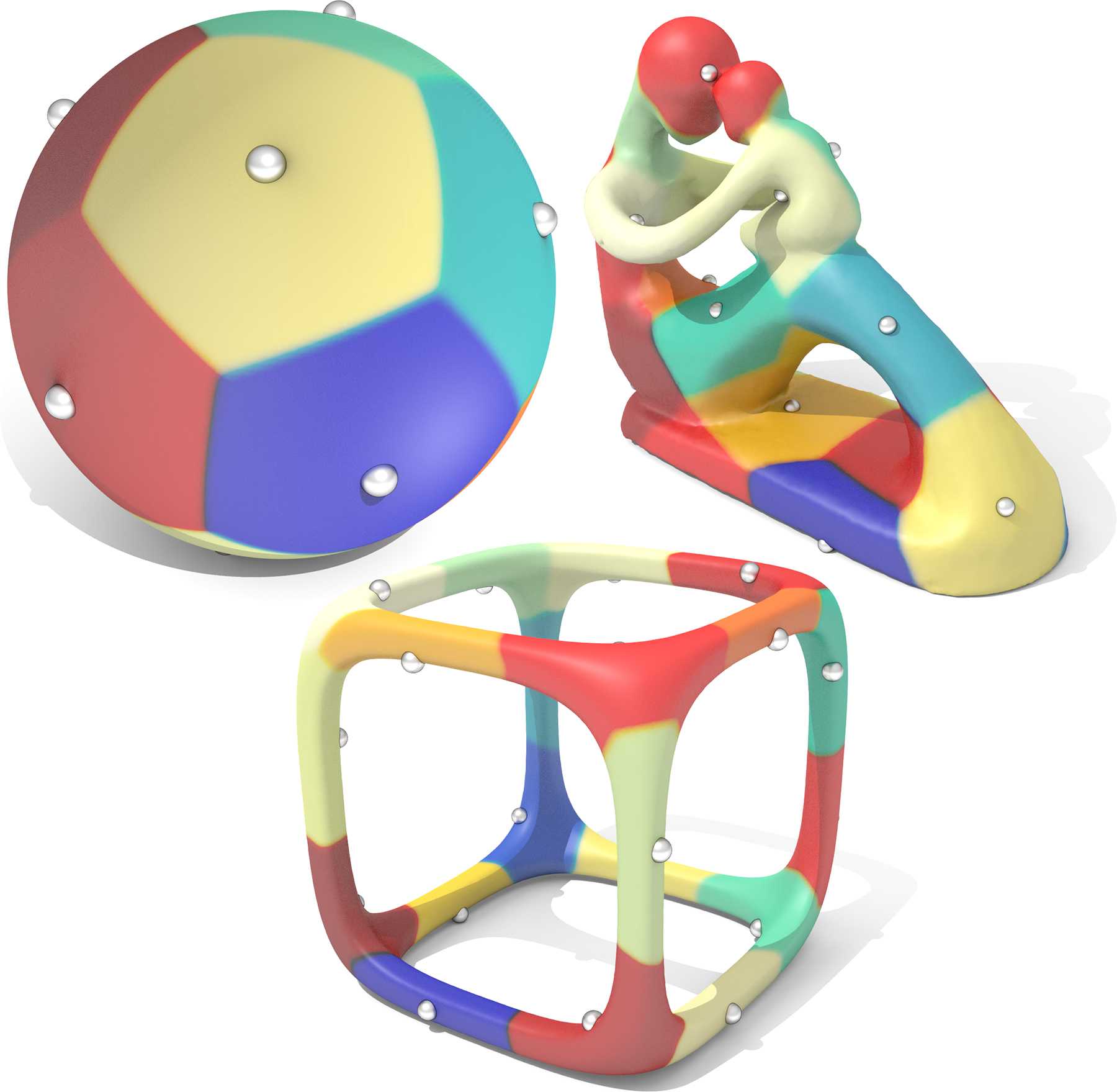}
   \caption{Fast Karcher means allow us to compute centroidal geodesic Voronoi tessellations with large, possibly multiply-connected cells.\label{fig:lloyds_algorithm}}
\end{wrapfigure}
A Voronoi diagram partitions a domain \(M\) into regions \(U_i \subset M\) comprised of those points closest to a given collection of \emph{sites} \(s_i \in M\).  In a centroidal Voronoi diagram, each site is located at the centroid of its associated region.  Our fast Karcher mean algorithm (\secref{KarcherMeans}) provides an effective way to compute \emph{geodesic centroidal Voronoi tessellations (GCVT)}, where the centroid is defined as the cell's Karcher mean.  Our method provides, to our knowledge, the first efficient approach for computing the true GCVT: using the Euclidean centroid (\ala{} CCVT~\cite{Du:2003:CCVT}) does not work well for large, curved cells; likewise, a diffusion-based centroid (\ala{} \citet{herholz2017diffusion}) can yield a diagram very different from the GCVT, especially when the Karcher mean is outside the cell.

To compute a GCVT, we simply apply Lloyd's algorithm, updating cell centers via the Karcher mean.  More specifically, we consider a distribution associated with each cell, defined via the scalar heat kernel \(k_t\) as
\begin{equation}
  \rho_i := \frac{k_t(s_i,\cdot)}{\sum_j k_t(s_j, \cdot)}
\end{equation}
for a small time \(t\).  This distribution is an indicator function for each cell \(U_i\); \(k_t\) is computed as usual (by solving a discrete heat equation).  We then apply \algref{karcher_mean} to move each site \(s_i\) to the center of the distribution $\rho_i$, and repeat until convergence. In practice, we find it is more efficient overall to take just a single step of the Karcher algorithm, even though it results in more Lloyd iterations.  We do not have to worry about topological issues like features separated by small extrinsic distances, and can handle multiply connected cells since we need only integrate the log map over each region \(U_i\).  Faster convergence might be achieved by replacing Lloyd's algorithm with a more sophisticated optimization strategy---\citet[Section 2]{Liu:2016:MDE} provides a nice discussion in the context of GCVT.

\subsection{Ordered Intrinsic Landmarks}
\label{sec:OrderedIntrinsicLandmarks}
\setlength{\columnsep}{1em}
\setlength{\intextsep}{0em}
\begin{wrapfigure}[20]{r}{134pt}
   \centering
   \includegraphics[width=\linewidth]{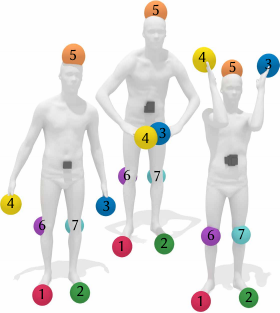}
   \caption{Our robust geometric median facilitates computation of consistently ordered landmarks, helping to avoid difficult matching problems. (Models reconstructed from scans in~\citet{Bogo:2014:FDE}.)\label{fig:landmarks}}
\end{wrapfigure}
An ongoing challenge in geometry processing is finding landmark points that provide correspondence between (near-)isometric surfaces.  A significant difficulty is not only finding geometrically salient points, but also finding a consistent \emph{ordering} for those points, \ie, which landmark on the first surface corresponds to which landmark on the second surface?  Trying to determine this ordering \aposteriori{} leads to hard combinatorial \emph{matching problems}; algorithms for efficiently computing such matchings are only just starting to be understood~\cite{Kezurer:2015:TRQ}.  The ability to reliably compute geometric centers provides new opportunities for generating landmarks that are consistently ordered \apriori{}.  We explore a simple strategy where we first compute the geometric medians \emph{of the surface itself}; in other words, we apply the algorithm described in \secref{KarcherMeans}, setting \(p=1\) and using a uniform density \(\rho = 1\) over the whole surface.  We then progressively add points via furthest point sampling, \ie, picking the point with greatest geodesic distance from our current set.  Since in general there may be more than one geometric median (\eg, on surfaces with symmetries) we sample a large set of initial guesses, and compute the furthest points relative to this \emph{set} (discarding the medians themselves, since we cannot easily distinguish their order).     Using the geometric median rather than the Karcher mean significantly reduces variability caused by values near the cut locus, which are effectively treated as outliers.  As is common in shape correspondence problems, we can add a weak extrinsic factor (\eg, the \(x\)-coordinate) to the density \(\rho\) in order to break symmetry.  Though we have so far considered only the most basic implementation, results are already promising (\figref{landmarks}); further refinement of such techniques would make for interesting future work.

\section{Limitations}
\label{sec:Limitations}

Potential challenges in applying our method stem mainly from two sources, namely (i) scalability of linear solvers, and (ii) numerical accuracy.  Since we solve standard Poisson-like systems, the scalability issue is no different than for many other problems in scientific computing---for instance, if meshes become too big to factor, one can switch to more scalable solvers (of the kind discussed in \secref{RelatedWork}).  In practice, however, we find that modern direct solvers~\cite{Chen:2008:ACS} provide an excellent solution up to millions of elements.  In terms of accuracy, poor mesh quality can lead to problems such as indefinite matrices and spurious flipped vectors (\figref{intrinsic_delaunay_maps}).  Using an intrinsic Delaunay triangulation helps significantly, though (as with any finite element method) meshes with too few elements or poorly distributed vertices can of course yield low-quality results.  The use of low-order elements limits us to linear convergence; as with the scalar heat method, spline and subdivision bases might yield higher order accuracy~\cite{deGoes:2016:SEC,Nguyen:2016:CFE}.  As discussed in \secref{BasicProperties} some basic properties of smooth parallel transport are not exactly preserved.  Finally, the applications explored in \secref{Applications} leave many open questions, such as preserving data along level sets, generalizing the definition of the log map for domains with boundary, and improving the robustness of landmark identification.

\section{Conclusion}
\label{sec:Conclusion}

Vector fields arising from parallel transport along shortest geodesics enable the implementation of many fundamental algorithms in geometry processing, yet surprisingly have received little prior attention.  We have presented a first method for efficiently and reliably computing such fields, though many questions remain.  For instance: how to properly formulate boundary conditions, how to improve accuracy (\eg, using more sophisticated finite element discretizations \cite{Arnold:2015:FEE}), how to improve practical efficiency (\eg, via parallelism and local computation), and what additional properties might be guaranteed (expanding on \secref{BasicProperties}).  Another interesting question is whether the method can be augmented to compute transport along non-geodesic curves, or along a given vector field (\ala{} \cite{Azencot:2015:DDV}).  Overall, given that the starting point (vector diffusion via the connection Laplacian) is quite unlike many traditional methods from computational geometry, we expect our method will inspire new ways of looking at old problems and lead to very different computational trade offs.  We are hopeful that the ease of implementation (just building and solving Laplace-like systems) will facilitate rapid adoption in real applications. 

\section*{Acknowledgements}

Thanks to Jooyoung Hahn for early discussions about signed distance computation, and Jim McCann for project feedback.  This work was supported by a Packard Fellowship, NSF Award 1717320, an NSF Graduate Research Fellowship, and gifts from Autodesk, Adobe, and Facebook.

\bibliographystyle{ACM-Reference-Format}
\bibliography{VectorHeatMethod}

\appendix

\section{Distance Gradient Discretization}
\label{app:DistanceGradientDiscretization}

For the log map (\secref{LogarithmicMap}), we need a discretization of the radial vector field \(R\), which in the smooth setting is the gradient of the distance function at \(x\), \ie, \(R|_y := \nabla d(x,y)\).  Away from the cut locus, \(R\) is a unit vector field tangent to the shortest geodesics emanating from \(x\).  To get an accurate discretization, we can therefore transport unit vectors in a small neighborhood around \(x\) to every other point---but must be careful about initial conditions.  Simply \emph{sampling} initial conditions onto vertices will not yield well-behaved solutions (\figref{gradient_discretization_comparison}, \figloc{center}); the dangers of intermingling sampling and finite-elements are well-known.  We instead take a finite element approach, leading to reliable and accurate initial conditions (\figref{gradient_discretization_comparison}, \figloc{right}).  We work in a flat Euclidean domain where we can use a single coordinate system for all tangent spaces; the resulting expressions also yield accurate results on curved domains due to the normalization by angle sums \(\Theta_i\) (\secref{IntrinsicTangentSpaces}), which effectively ``flattens out'' each vertex tangent space.

Consider a small circle \(\Ceps\) of radius \(\veps > 0\) centered around a source vertex \(i \in V\),  and let \(n\) denote its outward unit normal field.  For a point in the plane---or at the tip of a cone, as depicted in \secref{DiscreteSurface}---the normals \(n\) are exactly the gradient of geodesic distance along \(\Ceps\).  Just as one might treat a point source as a measure of unit mass supported on a point (\ie, a Dirac delta), we consider a measure of unit mass supported on \(\Ceps\), namely
\[
   \mu_{\veps} := \frac{1}{\veps^2}\mathcal{H}^1_{\Ceps},
\]
where \(\mathcal{H}^1_{\Ceps}\) is the Hausdorff measure on the circle.  Our initial conditions are then given by the vector-valued measure \(X_{\veps} := n \mu_{\veps}.\)

Now let $F:=\operatorname{span}\{\psi_v \st v\in V\}$ denote the finite element space of piecewise linear hat functions \(\psi_v\) at vertices \(v\).  To discretize a solution to the PDE \((\mathrm{id} + \ConnectionLaplacian)Y = X_{\veps}\)
in the space $F$, we solve
\[
   (\Msf + \Asf)\ysf = \xsf,
\]
where \(\Msf_{ij}\) is the mass matrix and \(\Asf\) is the stiffness matrix discretizing the connection Laplacian---in our case we use the matrices defined in \secref{DiscreteLaplaceOperators}.  We therefore just need to discretize the right-hand side, which is achieved by integrating each of the basis functions \(\psi_i\) with respect to the measure \(X_{\veps}\), \ie, by evaluating the integrals
\[
   \xsf_v := \int \psi_v~dX_\veps
\]
\setlength{\columnsep}{1em}
\setlength{\intextsep}{0em}
\begin{wrapfigure}[17]{r}{84pt}
   \includegraphics{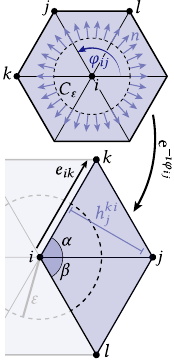}
\end{wrapfigure}
over the whole domain.  Since \(X_\veps\) is supported only on triangles containing the source vertex \(i\), and since each basis function \(\psi_v\) is supported only on the triangles containing \(v\), we need only evaluate this expression for immediate neighbors of \(i\).

For each neighbor \(j\), we work in a polar coordinate system \((r,\theta)\) with \(i\) at the origin and edge \(ij\) along the horizontal axis (see inset).  We treat points and vectors as complex numbers, using \(\ee\) to denote Euler's number and \(\imath\) to denote the imaginary unit (which applies a 90-degree counter-clockwise rotation); we use \(\langle \cdot, \cdot \rangle\) to denote the usual (real) inner product of vectors.  For brevity we will also let \(\alpha := \smash{\theta_i^{jk}}\) and \(\beta := \smash{\theta_i^{lj}}\).  In this coordinate system, we can express the unit normal as
\(n(\theta) := (\cos\theta,\sin\theta) = \ee^{\imath\theta}.\) At any point \(x\) in triangle \(ijk\), the piecewise linear hat function \(\psi_j\) can be expressed as
\[
   \psi_j(x) = \frac{1}{h_j^{ki}} \langle x, -\imath e_{ik}/\ell_{ik} \rangle = \frac{1}{2A_{ijk}} \langle x, -\imath e_{ik} \rangle
\]
where \(e_{ik}\) is the edge vector from \(i\) to \(k\), \(\ell_{ik} := |e_{ik}|\) is its length, \(\smash{h_j^{ki}}\) is the height of the triangle with apex \(j\) and base \(ki\), and \(A_{ijk}\) is its area.  In other words, to get a linear function with value 1 at \(j\) and value 0 at \(i\) and \(k\), we take the dot product with the unit vector orthogonal to \(e_{ik}\), and divide by height.  Integrating this function over the triangle \(ijk\) with respect to the measure \(X_\veps\) then yields
\begin{align*}
\int_{ijk} \psi_j dX_\veps &= \frac{1}{\veps^2} \int_0^{\alpha} \psi_j(\veps \ee^{\imath\theta}) n(\theta)\ \veps d\theta \\
&= \frac{\ell_{ik}}{2 A_{ijk}} \int_0^{\alpha} \langle \ee^{\imath\theta}, \ee^{\imath(\alpha-\pi/2)} \rangle \ee^{\imath\theta}\ d\theta,
\end{align*}
where we have used the relationship \(e_{ik} = \ell_{ik}\ee^{\imath\alpha}\).  Integrating this expression and repeating a nearly identical calculation for triangle \(jil\) we find that the overall integral of \(\psi_j\) is given by
\begin{equation}
   \label{eq:initial_data_neighbors}
   \boxed{
   \begin{array}{rcl}
     \tilde{\xsf}_j &=& \ell_{ik} \left( \alpha \sin\alpha, \sin\alpha     - \alpha\cos\alpha \right) / (4 A_{ijk})\ + \\
                    & & \ell_{il} \left( \beta  \sin\beta,  \beta\cos\beta - \sin\beta        \right) / (4 A_{jil}).
   \end{array}
}
\end{equation}
To get the final entry for the right-hand side at vertex \(j\), we rotate this value back into the tangent coordinate system of $j$:
\begin{equation}
   \label{eq:entry_rotation}
   \xsf_j = -\ee^{\imath\varphi_{ji}} \tilde{\xsf}_j.
\end{equation}
Since Equations~\ref{eq:initial_data_neighbors} and \ref{eq:entry_rotation} involve only lengths, angles, and areas, they can easily be evaluated on any triangle mesh.  Note that the angles \(\varphi_{ji}\) are exactly the same as those given in \eqref{outgoing_edge_angles}.

Importantly, the initial value \(\xsf_i\) at the source vertex \(i\) should \emph{not} necessarily be zero: this value does not represent the pointwise value of the initial data, but is rather just one of several nodes that determines the best piecewise linear approximation.  The calculation of this value is similar to the neighboring values except that we now take a sum over all triangles \(ijk\) containing vertex \(i\).  For each such triangle we will again construct a coordinate system with \(i\) at the origin and \(ij\) along the horizontal axis.  In this coordinate frame, the outward unit vector orthogonal to edge \(kj\) can be expressed as
\[
   v := \imath(\ell_{ij}\ee^{\imath0} - \ell_{ik}\ee^{\imath\alpha})/\ell_{jk},
\]
where \(\alpha := \theta_i^{jk}\).  Within triangle \(ijk\), the basis function \(\psi_i\) is then
\[
   \psi_i(x) = 1 - \sang{x,v}/h_i^{jk},
\]
and the integral over the triangle is
\begin{align*}
    \int_{ijk}\psi_i~dX_{\veps} & = \frac{1}{\veps^2}\int_{0}^{\alpha}\psi_i\left(\veps \ee^{\imath\theta}\right)n(\theta)~\veps d\theta \\
    & = \int_{0}^{\alpha}\ee^{\imath\theta}~d\theta - \frac{1}{2A_{ijk}}\int_{0}^{\alpha}\sang{\ee^{\imath\theta}, \imath(\ell_{ij} - \ell_{ik}\ee^{\imath\alpha}) }\ee^{\imath\theta}~d\theta.
\end{align*}
We can ignore the first term since over the whole circle it integrates to zero; the remaining term integrates to
\begin{equation}
   \boxed{
   \label{eq:initial_data_center}
   \tilde{\xsf}_{i,j} := \frac{1}{4A_{ijk}}\left[
      \begin{array}{c}
      -\sin\alpha(\ell_{ik}\alpha+\ell_{ij}\sin\alpha) \\
      \ell_{ij}(\cos\alpha\sin\alpha - \alpha) + \ell_{ik}(\alpha\cos\alpha - \sin\alpha) \\
      \end{array}
      \right].
}
\end{equation}
The initial value at the source is then given by the sum of these values, rotated back to the original coordinate system:
\[
  \xsf_i := \sum_{ijk \in F} \ee^{\imath\varphi_{ij}} \tilde{\xsf}_{i,j}.
\]
Again, these initial conditions can be easily computed on any triangle mesh. Notice that the initial conditions do not depend on the choice of radius \(\veps\).  To obtain the radial field \(R\) we now run \algref{vector_heat_method} with initial conditions \(\xsf\), but can simply normalize the resulting vectors rather than solving for magnitudes.  At the source vertex \(i\) we set \(R_i\) to zero.

\begin{figure}
   \centering
   \includegraphics{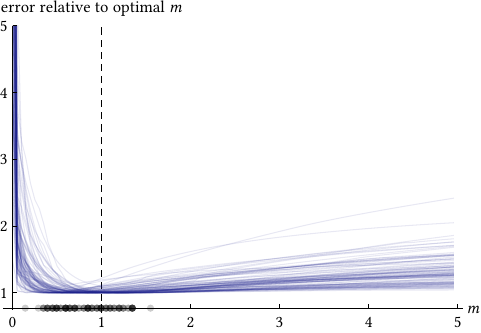}
   \caption{To determine a good choice for the parameter \(t\), we measured the error of the vector heat method for varying values of \(t = mh^2\), where \(h\) is the mean edge length.  Each curve represents a mesh from the data set of \protect{\citet{Myles:2014:RFG}}.  For each value of \(m\), the error is the mean error relative to the exact polyhedral solution over the whole surface.  Each curve is independently normalized by the error at its optimal \(m\) value, indicated by a black dot.  Overall, we find that \(t=h^2\) (\ie, \(m=1\)) is a reasonable choice across a wide variety of examples.\label{fig:ChoiceOfT}}
\end{figure}

\begin{figure}
   \centering
   \includegraphics[width=\columnwidth]{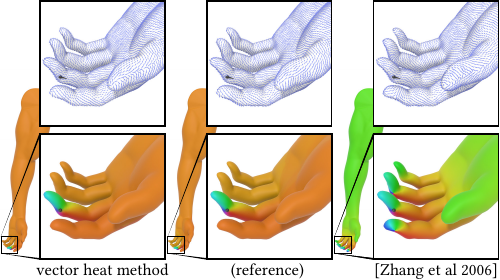}
   \caption{Other local approximations of parallel transport, such as the method of Zhang et al [2006], do not provide a good global approximation over the larger domain.  Here the field produced by the vector heat method closely matches the reference (exact polyhedral surface) allowing us to compute a globally accurate log map; the local approximation of Zhang et al deviates significantly away from the source.\label{fig:ArmComparison}}
\end{figure}

\end{document}